\documentclass[journal]{IEEEtran}
\usepackage{cite}

\ifCLASSINFOpdf
   \usepackage[pdftex]{graphicx}
\else
   \usepackage[dvips]{graphicx}
\fi
\usepackage{amsmath}
\DeclareMathOperator*{\argmax}{argmax}
\usepackage{amssymb}

\usepackage{algorithm}
\usepackage{algorithmic}

\usepackage{url}

\usepackage{booktabs}
\usepackage{multirow}
\usepackage[table]{xcolor}

\usepackage{color}

\usepackage[skip=2pt]{caption}

\usepackage{enumitem}

\usepackage[compact]{titlesec}

\usepackage{indentfirst}

\begin{document}
\title{Reliable Online Resource Allocation for Multi-User Semantic  Communications: A Constraint Bayesian Optimization Approach}

\author{
	Huawei~Hou,~\IEEEmembership{Student Member,~IEEE},
	Suzhi~Bi,~\IEEEmembership{Senior Member,~IEEE},
	Xian~Li,~\IEEEmembership{Senior Member,~IEEE},
	Haixia~Zhang,~\IEEEmembership{Senior Member,~IEEE}, and
	Zhi~Quan,~\IEEEmembership{Senior Member,~IEEE}
	
	\thanks{H.~Hou, S.~Bi, X.~Li and Z.~Quan are with the State Key Laboratory of Radio Frequency Heterogeneous Integration, and also with the College of Electronics and Information Engineering, Shenzhen University, Shenzhen, China 518060 (e-mail: 2350432016@email.szu.edu.cn, \{bsz,  xianli, zquan\}@szu.edu.cn). \textit{(Corresponding author: Suzhi Bi)}}
	\thanks{H.~Zhang is with the Institute of Intelligent
		Communication Technology, and also with the Shandong Key Laboratory of Intelligent Communication and Sensing-Computing Integration, Shandong University, Jinan 250061, Shandong, China (e-mail: haixia.zhang@sdu.edu.cn).}

}




\maketitle

\begin{abstract}
Semantic communication has been increasingly integrated into edge computing systems for reconstruction tasks, owing to its advantages in source compression, robustness to channel noise, and end-to-end reconstruction performance. However, the black-box nature of neural-network (NN)-based semantic codecs, together with the noisy transmission of semantic features, makes it difficult to allocate transmission resources and guarantee reconstruction quality for multiple users.
In this paper, we propose a reliable online resource allocation framework for a semantic-driven multi-user edge computing system, where multiple users encode source information into semantic features and offload reconstruction to an edge server. We formulate a multi-user resource optimization problem whose objective jointly accounts for system-wide reconstruction performance and transmission latency, under constraints that guarantee each user’s minimum reconstruction quality. To solve this problem, we develop a Bayesian optimization (BO)-based online algorithm that enables flexible control of the user-side semantic compression ratio (CR) and allocation of transmission rates.
The edge server jointly determines each user’s CR and transmission rate by exploiting Gaussian-process (GP) models that capture the relationship between reconstruction performance, signal-to-noise ratio (SNR), and CR, and by employing an acquisition function to select CRs that satisfy the performance quality constraints while maximizing the objective. Simulation results on high-resolution video-frame reconstruction datasets demonstrate that the proposed method selects near-optimal CRs via the GP surrogate and acquisition function, achieving a 98.03\% constraint-satisfaction rate and reducing transmission latency by more than 45\% compared with fixed-CR schemes.
\end{abstract}

\begin{IEEEkeywords}
Semantic Communication, edge computing, resource allocation, Bayesian optimization, Gaussian process.
\end{IEEEkeywords}

%
\IEEEpeerreviewmaketitle

\section{Introduction}
\subsection{Motivations and Contributions}
The recent development of deep learning-based semantic communication technology effectively reduces the communication resource required to offload intensive data in edge inference systems \cite{SC1, SC2, SC3, SC4}. Specifically, a semantic transmitter typically employs a deep learning-based joint source channel coding (DeepJSCC) encoder to extract semantic features from high-dimensional source data and transmits through noisy channels. Meanwhile, a semantic receiver directly maps the noisy features to reconstruct source information through a deep neural network (DNN) decoder, eliminating the need for bit-level error correction \cite{SC5,SC7,SC8}. Compared with conventional task offloading methods that transmit and process raw data, the semantic-based paradigm significantly improves both communication and task execution efficiency \cite{SC6}.
In practice, user mobility and channel fading effect cause rapid fluctuations in signal-to-noise ratio (SNR) as well as the reconstruction quality \cite{JSCC-f}. To maintain high reconstruction performance, dynamic compression adaptive to wireless channel conditions is crucial. That is, applying a larger compression ratio (CR) with higher data redundancy to enhance transmission reliability when channels deteriorate, and a smaller CR to improve transmission efficiency under favorable channel condition \cite{JSCC-A, SCA, Gong_TWC}. 
This challenge becomes particularly critical in multi-user scenarios, where limited communication resource is shared among the users. 
Such heterogeneous multi-user settings, comprising aerial and ground nodes served by edge servers, are envisioned as fundamental components of future integrated networks (e.g., space-air-ground-sea integrated networks), in which AI and learning techniques are increasingly employed across a wide range of network functions, from resource allocation to physical-layer security \cite{security1,security2}.
To achieve a trade-off between the reconstruction performance and transmission latency while guaranteeing the reconstruction quality requirement of each user, an online CR selection and resource allocation scheme is required. 

In semantic transmission, the inherent fixed output dimension of the DNN-based encoder prevents dynamic CR adjustment to achieve more efficient resource allocation \cite{DDJSCC}. 
To improve the flexibility of DNN-based semantic codecs, existing studies have proposed methods such as adaptive JSCC architectures or multi-rate training schemes to enable varying output dimensions \cite{JSCC-A2, predict_JSCC}. Nevertheless, for dynamic resource allocation in multi-user semantic edge inference systems, a fundamental challenge arises from the \textit{black-box} nature of the semantic pipeline: the reconstruction quality lacks an explicit analytical relationship with critical variables like SNR and CR, while being highly dependent on the input content \cite{content, image_cla}. This problem is exacerbated by the \textit{absence of source information} at the receiver, making real-time performance monitoring and control impossible \cite{SCAN}. These issues collectively make it exceptionally difficult to precisely allocate resources to meet the diverse reconstruction quality requirements of individual users in dynamic environments, leading to system reliability and fairness issues.
Consequently, existing data-driven approaches for this problem—whether based on learning a surrogate model \cite{JSCC_model}  or employing model-free optimization like deep reinforcement learning (DRL) \cite{RA1}—face respective limitations. Their effectiveness is inherently limited by the model generalization \cite{JIAO_IOT} capability or slow learning convergence. More recently, \textit{Bayesian optimization} (BO) has been explored as a sample-efficient alternative for black-box optimization \cite{rBO,cBO}, and applied in a multi-user edge video analytics system to maximize the average task performance \cite{Tang_BO, Xian_BO}. 
BO-based surrogate modeling can, in principle, capture the black-box relationship between task performance and SNR/CR, and thus facilitate single-user resource optimization. However, in a multi-user semantic communication system, the optimization objective includes achieving a tradeoff between reconstruction quality and latency, as well as ensuring the reconstruction performance requirements for each user. In this case, the implicit reconstruction quality–SNR/CR relationship appears in both the objective function and per-user reconstruction performance constraints, substantially complicating the optimization and analysis. \textcolor{black}{Existing variable-CR JSCC architectures \cite{JSCC-A}, \cite{JSCC-A2} are devised for point-to-point links, where a single transmitter adapts its output dimension according to its own channel state and no shared resource needs to be arbitrated among users. In multi-user scenarios, however, the CR decisions of different users are tightly coupled through the common transmission-rate budget: increasing the CR of one user to improve its reconstruction quality inevitably reduces the rate available to the other users and increases the system-wide latency. A per-link rate-adaptation rule therefore cannot, by itself, ensure that every user meets its own quality requirement under a shared budget.} Consequently, ensuring rigorous per-user performance guarantees in multi-user semantic communication systems remains an open challenge.

In this paper, we propose a reliable online resource allocation framework for multi-user edge inference systems. Our key insight is to formulate the black-box resource allocation problem with explicit and probabilistic performance constraints for each user, ensuring their reconstruction quality requirements are met with high probability under dynamic channel conditions. For this, we develop a constraint  BO-based algorithm dynamically adjusting the CR and transmission resource for each user based on real-time SNR measurements.
\textcolor{black}{Aligned with the reconstruction-oriented branch of semantic communication, in which the end task designates reconstruction quality as the semantic metric, our framework builds upon DeepJSCC \cite{predict_JSCC} as its architectural backbone. On top of this backbone, we contribute a resource-allocation layer that jointly adapts the CR and transmission resources across multiple users with per-user reconstruction quality guarantees.} 
The main contributions of this paper are summarized as follows:
\begin{itemize}
	  \item \textbf{Online Resource Allocation for Semantic Communication with Performance Guarantees:} We formulate a joint CR and transmission-rate allocation problem for multi-user semantic edge inference as a black-box stochastic optimization problem with per-user reconstruction quality constraints. The main challenge is that the receiver cannot directly assess the reconstruction quality in the absence of the source information. Moreover, this inability, combined with the lack of explicit analytical models to map reconstruction quality with SNR and CR, prevents the receiver from evaluating the impact of CR allocation on the reconstruction quality, making it difficult to achieve per-user performance guarantee.
	\item \textbf{Oracle-Network-Based Online Performance Feedback Mechanism:} To address the lack of source information during online transmission, we design a feedforward oracle network that approximates the offline-trained edge inference model and predicts reconstruction quality as a function of the semantic features, SNR, and CR. This oracle deployed on the transmitter provides low-cost, real-time performance feedback under the given CR, assisting the centralized online resource allocation without additional information.
	\item \textbf{Constraint Bayesian Optimization Algorithm for Reliable Semantic Transmission}: We develop a constraint BO algorithm that uses Gaussian processes (GPs) to jointly model the reconstruction quality and constraint satisfaction over the SNR-CR space. Based on these GP models, we construct a constraint-aware acquisition function for online CR selection and transmission-rate allocation, which simultaneously seeks to maximize a system-level trade-off between weighted sum reconstruction performance and total latency, as well as enforcing the probabilistic constraint of the users' reconstruction quality requirements.

\end{itemize}

We evaluate the performance of the proposed method on multiple high-resolution video datasets collected from real-world scenarios,  including autonomous driving, aerial surveillance, and UAV drone. Results demonstrate that our method achieves a 98.03\% satisfaction rate for user performance constraints and reduces the transmission latency by over 45\% compared to non-adaptive fixed-CR strategies.

\subsection{Related Works}
\begin{figure*}[!t]
	\centering
	\includegraphics[width=6.5in]{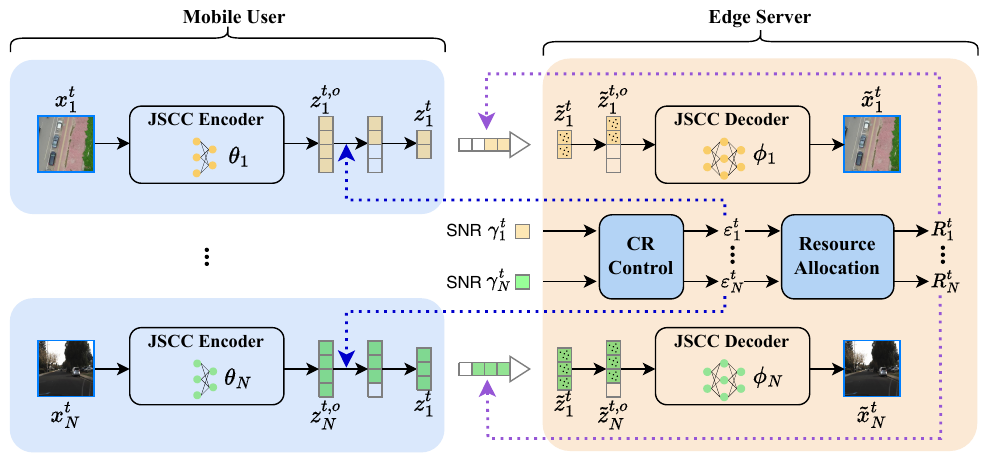}
	\caption{Block diagram of the proposed multi-user semantic edge reconstruction system with online resource allocation.}
	\label{fig_2}
\end{figure*}

\begin{figure}[!t]
	\centering
	\includegraphics[width=3.5in]{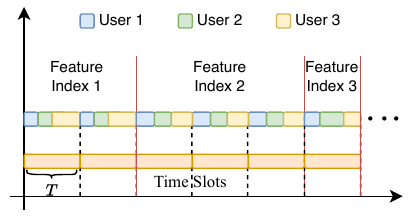}
	\caption{An illustration of TDMA-based transmission scheme.
	}
	\label{fig_tdma}
\end{figure}

Due to the black-box nature of DNN-based semantic codecs, it is generally intractable to derive an explicit analytical mapping between inference performance and key resource optimization variables. To address this issue, existing approaches mainly follow two directions: surrogate modeling and model-free reinforcement learning.

\subsubsection{Surrogate model-based methods} One line of research constructs parametric surrogate models to approximate the implicit performance relationship, and then optimizes system objectives (e.g., power, energy, bandwidth) based on the designed models. For instance, \cite{JSCC_model} proposes an AlphaBeta-Gamma (ABG) function to model the relationship between end-to-end performance metrics and SNR, which assists in guaranteeing quality of service (QoS) and designing optimal power allocation schemes. \cite{JSCC_LR} adopts a generalized logistic function to approximate the relation between the average SSIM and SNR under a fixed CR. \cite{JSCC_EXP} employs an exponential function to estimate the mapping between task performance and CR, while \cite{JSCC_Poly} establishes a quadratic polynomial model relating task accuracy to SNR to reduce optimization complexity. A common characteristic of these methods is the use of pre-specified functional forms with parameters derived from offline curve fitting. While effective in static settings, the fixed model structure could be less adaptable to distribution shifts, and model rebuilding might be needed when the system environment or tasks undergo significant changes.

\subsubsection{Reinforcement learning-based methods} Alternatively, another line of work formulates resource allocation as a Markov Decision Process (MDP) and applies model-free DRL to learn allocation policies through interaction with the environment. For example, \cite{RA1} employs DRL to jointly optimize CR, bandwidth, and transmit power under performance constraints, while \cite{RA2} incorporates channel conditions into DRL-based spectral efficiency optimization, enabling user-specific allocation of channels and CRs. \cite{DRL_CR} develops a DRL-based CR selection strategy conditioned on SNR to reduce transmitted data volume while maintaining task accuracy. For semantic video transmission, \cite{SC_video} introduces a reinforcement learning-based approach to explicitly balance transmission rate and task accuracy. Although DRL avoids the need for explicit performance modeling, it often suffers from slow convergence and training instability, particularly in dynamic multi-user scenarios, limiting its suitability for practical real-time systems requiring rapid and robust adaptation.

The rest of the paper is organized as follows. Section \ref{system model} presents the system model and formulates the optimization problem.
Section \ref{JSCC} describes the architecture and training process of the semantic encoder and decoder.
Section \ref{sec:cBO} introduces the proposed BO-based CR selection and bandwidth allocation framework. Section \ref{result} evaluates the performance of the proposed scheme through comparisons with representative benchmarks. Finally, Section \ref{clu} concludes the paper.

\begin{table}[!t]
	\renewcommand{\arraystretch}{1.3} 
	\caption{{Nomenclature}}
	\label{table_nomenclature}
	\centering
	\begin{tabular}{l p{7cm}}
		\hline
		\hline
		Symbol & Definition \\
		\hline
		$N$ & Total number of users \\
		$t$ & Index of time step, data, or feature \\ 
		$\boldsymbol{x}_n^t$ & Source data for the $n$th user at index $t$ \\
		$f_{\boldsymbol{\theta}_n}$ & JSCC encoder for the $n$th user, parameterized by $\boldsymbol{\theta}_n$ \\
		$\boldsymbol{z}_n^t$ & Extracted semantic feature for the $n$th user at index $t$ \\
		$\gamma_n^t$ & Instantaneous SNR for the $n$th user at index $t$ \\
		$\varepsilon_n^t$ & Allocated compression ratio (CR) for the $n$th user at index $t$ \\
		$R$ & Transmission rate \\
		$B$ & Number of quantization bits per complex symbol \\
		$h_n^t$ & Channel coefficient for the $n$th user at index $t$ \\
		$\epsilon_n^t$ & Gaussian channel noise vector for the $n$th user at index $t$ \\
		$f_{\boldsymbol{\phi}_n}$ & JSCC decoder for the $n$th user, parameterized by $\boldsymbol{\phi}_n$ \\
		$Q_n^t$ & Reconstruction quality metric for the $n$th user at index $t$ \\
		$\alpha$ & Weighting factor balancing reconstruction quality and latency \\
		$f_{\boldsymbol{\omega}_n}$ & Oracle network for the $n$th user, parameterized by $\boldsymbol{\omega}_n$ \\
		$\boldsymbol{\mathcal{D}}_n^{t-1}$ & Aggregated observations for the $n$th user up to index $t-1$ \\
		$t_u$ & Number of recent observations used for the GP model update \\
		$t_l$ & Update interval of the GP model \\
		\hline
		\hline
	\end{tabular}
\end{table}

\section{System Model and Problem Formulation} \label{system model}
As illustrated in Fig. \ref{fig_2}, we consider a multi-user semantic semantic edge reconstruction system comprising multiple heterogeneous mobile users, each equipped with an Adaptive JSCC module (detailed in Section \ref{JSCC}), which transmit compressed semantic features through noisy channels to the edge server. The server processes the received noisy features to reconstruct source data. Besides, it performs real-time CR control and transmission rate allocation of all users (detailed in Section \ref{sec:cBO}).

\subsection{Semantic Feature Transmission}
We consider $N$ users transmitting semantic features to the edge server in a TDMA (time division multiple access) manner \cite{shao2, SC_TDMA}, where time is divided into equal time slots each of duration $T$. We further assume that the users are executing independent inference tasks.  Consequently, the edge server processes the received semantic features from each user individually and in parallel. In the transmission of the $t$th feature of all users, we denote the observed $L_n^s$-dimensional source data of the $n$th user as $\boldsymbol{x}_n^t \in \mathbb{R}^{L_n^s}$, and the encoded complex semantic feature $\boldsymbol{z}_n^t \in \mathbb{C}^{L_n^t}$ as
\begin{equation} \label{se}
	\boldsymbol{z}_n^t = f_{\boldsymbol{\theta}_n}(\boldsymbol{x}_n^t, \gamma_n^t, \varepsilon_n^t), 
\end{equation}
where the $f_{\boldsymbol{\theta}_n}$ denotes the JSCC encoder parameterized by $\boldsymbol{\theta}_n$ \cite{SC4}. Specifically, we assume the channel remains unchanged during the transmission of the $t$th feature, and denote $\gamma_n^t$ as the instantaneous channel SNR defined later in (\ref{lb_snr}). $\varepsilon_n^t=L_n^t/L_n^s$ represents the CR with $\varepsilon_n^t \in [\varepsilon_n^{\min}, \varepsilon_n^{\max}]$. 
To ensure integer output dimension, in practice, we round $L_n^t$ as 
\begin{equation} \label{lnt}
	L_n^t = \left\lceil \varepsilon_n^t L_n^s \right\rceil, 
\end{equation}
where $\lceil x \rceil$ denotes the smallest integer larger than or equal to $x$. 
Notice that each vector $\boldsymbol{z}_n^t \in \mathbb{C}^{L_n^t}$ consists of $L_n^t$ complex numbers. 
To align with modern digital communication systems and enable bit-level resource allocation, we quantize the real and imaginary parts of continuous semantic features $\boldsymbol{z}_n^t$ into discrete representations \cite{DeepJSCC_Q}. Assuming each complex dimension is quantized using $B$ bits, the total data payload generated by the $n$th user for the $t$th feature is $B \cdot L_n^t$ bits. These bit streams are then mapped into digital modulation symbols (e.g., QAM) and transmitted over the physical wireless channels.

We model the end-to-end transmission pipeline, encompassing quantization, digital modulation, physical channel distortion, and demodulation/dequantization—as an \textit{equivalent baseband channel} in the semantic feature space \cite{DeepJSCC_Q2}.  Accordingly, the received feature vector at the edge server is 
\begin{equation} \label{eq_channel}
	\tilde{\boldsymbol{z}}_n^t= h_n^t \cdot  \boldsymbol{z}_n^t + \boldsymbol{\epsilon}_n^t, 
\end{equation}
where $h_n^t$ represents the effective channel fading coefficient, and $\boldsymbol{\epsilon}_n^t \sim \mathcal{CN}(0, \sigma_n^2 \boldsymbol{I})$  is an aggregate noise term.\footnote{To address the non-differentiability of quantization and modulation during end-to-end training of the codecs, we can apply a differentiable surrogate model to equivalently represent the quantization and modulation procedure  \cite{xian_tvt} to facilitate gradient backpropagation.} In this paper, we consider a block fading channel, where $h_n^t$ remains constant during the transmission of one semantic feature vector and may vary independently across different rounds of transmissions.

Before transmission, the power of $\boldsymbol{z}_n^t$ is normalized to satisfy 
\begin{equation}
	\frac{\mathbb{E}[\|\boldsymbol{z}_n^t\|_2^2]}{L_n^t} = P_{n},
\end{equation}
where $P_{n}$ is the power constraint. We then define the instantaneous channel SNR as
\begin{equation} \label{lb_snr}
	\gamma_n^t = \frac{|h_n^t|^2}{\sigma_n^2}.
\end{equation}
Without loss of generality, we set $h^t_n\sim \mathcal{CN}(0,1)$ and $P_n=1$ for all $n$.
Suppose that the total transmission rate of the system is $R$ bits/slot. By allocating $R_n^t$ transmission rate to user $n$, we have 
\begin{equation}
		\sum_{n=1}^{N} R_n^t \leq R, \quad \forall t.
\end{equation}
Accordingly, the transmission time for user $n$'s features is
\begin{equation}
	T_n^t = \frac{BT L_n^t}{R_n^t}.
\end{equation}
Since the tasks are independent, the system efficiency is determined by the average response time across all users. Hence, we denote the average transmission latency across all users as
\begin{equation}
	\frac{1}{N} \sum_{n=1}^N T_n^t = \frac{BT}{N} \sum_{n=1}^N \frac{L_n^t}{R_n^t} = \frac{BT}{N} \sum_{n=1}^N \frac{\varepsilon_n^t L_n^s}{R_n^t}.
\end{equation}

Upon receiving the noisy semantic feature $\tilde{\boldsymbol{z}}_n^t$, the edge server performs source information reconstruction
\begin{equation}
	\tilde{\boldsymbol{x}}_n^t=f_{\boldsymbol{\phi}_n} (\tilde{\boldsymbol{z}}_n^t, \gamma_n^t), 
\end{equation}
where $f_{\boldsymbol{\phi}_n}$ is the JSCC decoder parameterized by $\boldsymbol{\phi}_n$. 
We denote $Q_n^t(\tilde{\boldsymbol{x}}_n^t, {\boldsymbol{x}}_n^t)$ as the performance metric for evaluating the reconstruction quality, e.g., peak signal-to-noise ratio (PSNR) for image reconstruction.

\subsection{Problem Formulation}
The core challenge lies in optimizing the trade-off between reconstruction quality and transmission latency under dynamic channel conditions. This requires joint adaptation of CRs $\boldsymbol{\varepsilon}^t$ and rate allocation $\boldsymbol{R}^t$. Intuitively, higher CRs ($\varepsilon_n^t$) improve reconstruction quality $Q_n^t$ but increase transmission latency, and vice versa.  
Accordingly, we formulate the following optimization problem:
\begin{subequations}
	\begin{align}
		\text{(P1):}	\quad  \max_{\boldsymbol{\varepsilon}^t, \boldsymbol{R}^t} & \  \sum_{n=1}^{N} \left(  Q_n^t  - \alpha \cdot \frac{\varepsilon_n^t L_n^s B T}{NR_n^t} \right),   \label{P1}  \\
		\text { s.t. } & Q_n^t  \ge Q_n^{\min}, \quad \forall n, \forall t,  \label{P1_c1} \\  
		&  \sum_{n=1}^{N}  R_n^t  \leq R,  \quad \forall t,  \label{P1_c2} \\
		& R_n^t \ge 0, \quad \forall n,t, \label{P1_c3} \\
		&   \varepsilon_n^{\min} \le  \varepsilon_n^t \le   \varepsilon_n^{\max}, \quad \forall n,t, \label{P1_c4} 
	\end{align}
\end{subequations}
where (\ref{P1}) is the optimization target function, which aims to maximize the aggregate quality-latency utility of the system with respect to the average performance for independent multi-user access. $\alpha > 0$ is a weighting factor that balances the reconstruction quality and latency. The constraint (\ref{P1_c1}) is the minimal reconstruction quality requirement of each user. (\ref{P1_c2}) is the total transmission rate budget constraint. 

The optimization problem (\ref{P1}) presents twofold fundamental challenges. 
First, during online semantic transmission, as the source data $\boldsymbol{x}_n^t$ is unavailable at the edge server, the receiver cannot evaluate the reconstruction quality and issue effective control decisions to optimize (\ref{P1}). 
Second, performance metric $Q_n^t$ is a black-box function with respect to $(\boldsymbol{z}_n^t, \varepsilon_n^t, \gamma_n^t)$ without explicit analytical form. It is very hard to optimize the objective and even harder to satisfy the constraint (\ref{P1_c1}). 
To overcome these challenges, we develop an oracle network in Section \ref{JSCC} to establish a mapping between $Q_n^t$ and $(\boldsymbol{z}_n^t, \varepsilon_n^t, \gamma_n^t)$ through offline learning. Then, in Section \ref{sec:cBO}, we propose a BO-based online black-box optimization method to solve (P1) under real-time channel variations.

\section{Adaptive JSCC and Oracle Network}  \label{JSCC}
In this section, we first describe the design and training procedures of the CR-adaptive JSCC framework that can flexibly adjust the dimension of the transmitted feature. Then, we introduce a feedforward oracle network deployed at the user device to address the problem that the edge receiver cannot evaluate the reconstruction quality. For simplicity of exposition, we omit the time index $t$ during the offline training process of the JSCC codecs and the oracle network.

\subsection{JSCC Encoder and Decoder}
The semantic encoder of the $n$th user is implemented as a NN parameterized by $\boldsymbol{\theta}_n$. To achieve flexible CR adaptation, we employ a masking mechanism on the fixed-dimensional output feature \cite{predict_JSCC}. In particular, we denote the fixed-dimension output of the encoder as
\begin{equation} \label{zo}
	\boldsymbol{z}_n^o =  f_{\boldsymbol{\theta}_n}(\boldsymbol{x}_n, \gamma_n), 
\end{equation}
where $\boldsymbol{z}_n^o \in \mathbb{C}^{L_n^{\max}}$ denotes the full-dimensional feature with $L_n^{\max} = L_n^C \times L_n^W \times L_n^H$. Here, $L_n^C$, $L_n^W$, and $L_n^H$ are channel depth, width, and height dimensions of the convolutional feature map, respectively.\footnote{In this context, ``channel" refers to the feature channels in deep learning architectures, representing distinct filter responses (e.g., a 64-channel feature map indicates 64 learned convolutional kernels). This differs from the communication channel terminology used elsewhere in the paper.}
Given the CR $\varepsilon_n$, the masked feature $\bar{\boldsymbol{z}}_n$ is obtained through:
\begin{equation} \label{mask}
	\bar{\boldsymbol{z}}_n = \boldsymbol{z}_n^o \odot \boldsymbol{\delta}_n, 
\end{equation}
where $\odot$ denotes element-wise multiplication, and $\boldsymbol{\delta}_n$ is a binary mask vector defined as
\begin{equation}
	\boldsymbol{\delta}_n = \left\{\begin{array}{ll}
		1, & \text { if } i \leq L_n \\
		0, & \text { otherwise}.
	\end{array}\right.
\end{equation}
Here, $L_n = \lceil \varepsilon_n L_n^s \rceil$ denotes the active feature length. This yields $\bar{\boldsymbol{z}}_n = [\boldsymbol{z}_n; \boldsymbol{0}]$, where $\boldsymbol{z}_n \in \mathbb{C}^{L_n}$ contains the transmitted features and $\boldsymbol{0} \in \mathbb{C}^{L_n^{\max}-L_n}$ is a zero-padding vector.
\textcolor{black}{The mask-based training allows the DeepJSCC codec to perform robustly at supported CRs, effectively adapting its feature representation to different compression levels. More importantly, the mask reduces the CR to a single scalar for the centralized controller at the edge server, resulting in an a priori known transmission payload $B\lceil \varepsilon_n L_n^s \rceil$. This renders resource allocation subproblem (P2) well posed, which in turn admits the closed-form solution in \eqref{tsa}. Since the mask is fully determined by $\varepsilon_n$ issued by the edge server, no mask pattern needs to be conveyed to the decoder as side information for each frame and each user.}

At the receiver, the noisy features $\tilde{\boldsymbol{z}}_n$ are firstly zero-padded to reconstruct the full-dimensional input $\tilde{\boldsymbol{z}}_n^o = [\tilde{\boldsymbol{z}}_n; \boldsymbol{0}]$.
Then, the source data reconstruction is performed by
\begin{equation} \label{xn}
	\tilde{\boldsymbol{x}}_n = f_{\boldsymbol{\phi}_n}(\tilde{\boldsymbol{z}}_n^o, \gamma_n),
\end{equation}
where $f_{\boldsymbol{\phi}_n}$ denotes the JSCC decoder parameterized by $\boldsymbol{\phi}_n$. The reconstruction quality is evaluated by $Q_n(\boldsymbol{x}_n, \tilde{\boldsymbol{x}}_n)$.

\subsection{Oracle Network}
Because the ground-truth source data $\boldsymbol{x}_n$ is unavailable at the receiver in real-time inference, the receiver is unable to compute  $Q_n$ and optimize (P1) accordingly. 
To address this issue, we deploy a feedforward oracle network \cite{predict_JSCC} at the transmitter side to predict the reconstruction quality. Specifically, we denote the predicted $Q_n$ as 
\begin{equation} \label{an}
	\tilde{Q}_n = f_{\boldsymbol{\omega}_n}(\boldsymbol{\mu}_n^o, \boldsymbol{\sigma}_n^o, \gamma_n, \varepsilon_n),
\end{equation} 
where $f_{\boldsymbol{\omega}_n}$ represents the oracle network with trainable parameters $\boldsymbol{\omega}_n$. 
$\boldsymbol{\mu}_n^o = [\mu_n^1,...,\mu_n^{L_n^C}]^\top$ and $\boldsymbol{\sigma}_n^o = [\sigma_n^1,...,\sigma_n^{L_n^C}]^\top$ are the aggregated channel-wise statistics of $\boldsymbol{z}_n^o$ , where $\mu_n^c \in \boldsymbol{\mu}_n^o $ and $\sigma_n^c \in \boldsymbol{\sigma}_n^o$ denote the mean and standard deviation of the $c$th feature channel, i.e., 
\begin{equation}
		\mu_n^c = \frac{1}{L_n^W \times L_n^H} \sum_{i=1}^{L_n^W} \sum_{j=1}^{L_n^H} z_n^{o,(c)}[i,j], \ \  c \in \{1,...,L_n^C\}, 
\end{equation}
\begin{equation}
		\sigma_n^c = \sqrt{\frac{1}{L_n^W \times L_n^H} \sum_{i,j} |z_n^{o,(c)}[i,j] - \mu_n^c|^2}, \ \  c \in \{1,...,L_n^C\}.
\end{equation}

\begin{algorithm}[t]
	\caption{Multi-user Adaptive JSCC Training}
	\label{jscc_training}
	\renewcommand{\algorithmicrequire}{\textbf{Input:}}
	\renewcommand{\algorithmicensure}{\textbf{Output:}}
	\begin{algorithmic}[1]
		\REQUIRE Training datasets $\mathcal{D}^{\text{train}} = [\mathcal{D}_1^{\text{train}}, \cdots, \mathcal{D}_N^{\text{train}}]$
		\ENSURE Trained Adaptive JSCC encoder and decoder $\{[\boldsymbol{\theta}, \boldsymbol{\phi}]\} = \{[\boldsymbol{\theta}_1, \boldsymbol{\phi}_1], \cdots, [\boldsymbol{\theta}_N, \boldsymbol{\phi}_N]\}$
		
		\STATE Initialize model parameters $\{[\boldsymbol{\theta}, \boldsymbol{\phi}]\}$
		\FOR{$n=1$ to $N$}
		\REPEAT
		\STATE Sample data batch $\{\boldsymbol{x}_{n,b}\}_{b=1}^{B_j}$ from $\mathcal{D}_n^{\text{train}}$
		\STATE Sample CR batch $\{\varepsilon_{n,b}\}_{b=1}^{B_j} \sim \mathcal{U}[\varepsilon_n^{\min}, \varepsilon_n^{\max}]$
		\STATE Sample SNR batch $\{\gamma_{n,b}\}_{b=1}^{B_j} \sim \mathcal{U}[\gamma_n^{\min}, \gamma_n^{\max}]$
		\FOR{$b=1$ to $B_j$}
		\STATE Extract features $\boldsymbol{z}_{n,b}^o $ \hfill $\triangleright$ Eq. (\ref{zo})
		\STATE Apply mask: $\boldsymbol{z}_{n,b} = \boldsymbol{z}_{n,b}^o \odot \boldsymbol{\delta}_{n,b}$ \hfill $\triangleright$ Eq. (\ref{mask})
		\STATE Transmit semantic feature
		\STATE Pad received signal: $\tilde{\boldsymbol{z}}_{n,b}^o = [\tilde{\boldsymbol{z}}_{n,b}; \boldsymbol{0}]$ 
		\STATE Recover source $\tilde{\boldsymbol{x}}_{n,b} $ \hfill $\triangleright$ Eq. (\ref{xn})
		\STATE Compute $\mathcal{L}_{\boldsymbol{\theta}_n, \boldsymbol{\phi}_n} (\boldsymbol{x}_n,  \tilde{\boldsymbol{x}}_n, \varepsilon_n, \gamma_n )$ \hfill $\triangleright$ Eq. (\ref{loss_jscc})
		\ENDFOR
		\STATE Update $\boldsymbol{\theta}_n, \boldsymbol{\phi}_n$ via $\nabla_{\boldsymbol{\theta}_n, \boldsymbol{\phi}_n}\mathcal{L}_{\boldsymbol{\theta}_n, \boldsymbol{\phi}_n} (\boldsymbol{x}_n,  \tilde{\boldsymbol{x}}_n, \varepsilon_n, \gamma_n )$
		\UNTIL Convergence condition is met.	
		\ENDFOR
		\FOR{$n=1$ to $N$}
		\REPEAT
		\STATE Sample batches $\{\boldsymbol{x}_{n,b}\}_{b=1}^{B_o}$, $\{\varepsilon_{n,b}\}_{b=1}^{B_o}$ and $\{\gamma_{n,b}\}_{b=1}^{B_o}$
		\FOR{$b=1$ to $B_o$}
		\STATE Extract features $\boldsymbol{z}_{n,b}^o$ using pre-trained $f_{\boldsymbol{\theta}_n}$
		\STATE Compute channel-wise statistics $\boldsymbol{\mu}_{n,b}^o $ and  $\boldsymbol{\sigma}_{n,b}^o$
		\STATE Predict metric $\tilde{Q}_{n,b}$ \hfill $\triangleright$ Eq. (\ref{an})
		\STATE Recover source $\tilde{\boldsymbol{x}}_{n,b} $ using pre-trained $f_{\boldsymbol{\phi}_n}$
		\STATE Compute $\mathcal{L}_{\boldsymbol{\omega}_n} (\boldsymbol{z}_n,  \varepsilon_n, \gamma_n )$ \hfill $\triangleright$ Eq. (\ref{lo})
		\ENDFOR
		\STATE Update $\boldsymbol{\omega}_n$ via $\nabla_{\boldsymbol{\omega}_n}\mathcal{L}_{\boldsymbol{\omega}_n} (\boldsymbol{z}_n,  \varepsilon_n, \gamma_n )$
		\UNTIL Convergence condition is met.	
		\ENDFOR
	\end{algorithmic}
\end{algorithm}

\subsection{Model Training}
The JSCC codecs are trained in an end-to-end manner using the mean squared error (MSE) loss
\begin{equation} \label{loss_jscc}
	\mathcal{L}_{\boldsymbol{\theta}_n, \boldsymbol{\phi}_n} (\boldsymbol{x}_n,  \tilde{\boldsymbol{x}}_n, \varepsilon_n, \gamma_n ) = \frac{1}{L_n} \left \| \boldsymbol{x}_n- \tilde{\boldsymbol{x}}_n\right \| ^2. 
\end{equation}
Subsequently, the oracle network is trained separately using the MSE between predicted and actual quality metrics
\begin{equation} \label{lo}
	\mathcal{L}_{\boldsymbol{\omega}_n} (\boldsymbol{\mu}_n^o, \boldsymbol{\sigma}_n^o,  \varepsilon_n, \gamma_n ) = \| Q_n-\tilde{Q}_n \| ^2. 
\end{equation}
During training, both CR values $\varepsilon_n$ and SNR values $\gamma_n$ are uniformly sampled from their respective ranges $[\varepsilon_n^{\min}, \varepsilon_n^{\max}]$ and $[\gamma_n^{\min}, \gamma_n^{\max}]$ to ensure robust operation across diverse conditions. The training procedures of the JSCC codecs and oracle network are summarized in Algorithm \ref{jscc_training}. After this offline training, we fix the parameters of the DeepJSCC codecs $\{\boldsymbol{\theta}_n, \boldsymbol{\phi}_n\}_{n=1}^N$ and the oracle network $\{\boldsymbol{\psi}_n\}_{n=1}^N$. At the online stage, the CR decisions rely on the real-time updating and predicting of the GP model, detailed in Section \ref{sec:cBO}. The ground truth PSNR $Q_n^t$ is not needed during online operations.

\section{Proposed BO-based Resource Allocation Method} \label{sec:cBO}
In this section, we propose an efficient algorithm to solve (P1). Specifically, we first decompose (P1) into two subproblems that optimize the transmission rate allocation and CR, respectively. Then, we derive the optimal transmission rate allocation solution and reduce (P1) to a simplified form that optimizes only CR. Accordingly, we propose a constrained BO-based solution and discuss the computational complexity. 

As the reconstruction quality metric $Q_n^t$ is unobservable at the edge server due to the lack of source information, we leverage the oracle network introduced in Section \ref{JSCC} to predict $\tilde{Q}_n^t$ at the transmitter side, which serves as a surrogate for $Q_n^t$ in the optimization. Based on this prediction, we propose an online algorithm framework to solve (P1) for each index $t$.
We describe the procedure step-by-step as follows:
\begin{itemize}
	\item \textbf{Step 1: CR selection via BO: }The edge server obtains SNRs $\boldsymbol{\gamma}^t$ by estimating the uplink channels of all users, then employs the proposed BO-based method to determine the CR for each user, i.e., $\varepsilon_n^t \in [\varepsilon_n^{\min}, \varepsilon_n^{\max}]$.
	\item \textbf{Step 2: transmission rate allocation: }Based on $\boldsymbol{\varepsilon}^t$, the edge server allocates the optimal transmission resources $\boldsymbol{R}^t$ by solving the rate allocation subproblem of (P1).

\end{itemize}
This workflow repeats from Step 1 for transmitting a new set of features, producing new resource allocation decisions under varying channel conditions and perceived contents. 
In the following subsections, we will first investigate the optimal transmission rate allocation scheme under given CRs in Section \ref{oba}. Then, Sections \ref{gp} elaborate on the proposed BO-based method, including GP modeling, acquisition function, and complexity analysis.

\subsection{Customized Optimal Transmission Rate Allocation} \label{oba}
Suppose that a feasible $\boldsymbol{\varepsilon}^t$ is given. 
As the $Q^t_n$ in (\ref{P1}) is related only to $\varepsilon^t_n$, we can remove the corresponding terms in the objective and constraints of (P1), which reduces the problem to the following transmission rate allocation problem:
\begin{subequations}
	\begin{align}
		\text{(P2):}	\quad  	\min_{\boldsymbol{R}^t} & \  \frac{BT}{N}\sum_{n=1}^{N} \frac{\varepsilon_n^t L_n^s }{R_n^t},   \label{p2} \\
		\text { s.t. } & \sum_{n=1}^{N}  R_n^t  \leq R,  \forall n,   \label{p2_c1}   \\
		& R_n^t \ge 0, \quad \forall n.  \label{p2_c2}
	\end{align}
\end{subequations}

It is evident that (P2) is a convex optimization and the constraint (21b) is tight at the optimum. By introducing a Lagrangian multiplier $\lambda \geq 0$ to (21b), we have 
\begin{equation}
	\mathcal{L}(\boldsymbol{R}^t, \lambda) = \sum_{n=1}^N \frac{\varepsilon_n^t L_n^s}{R_n^t} + \lambda \left( \sum_{n=1}^N R_n^t - R \right).
\end{equation}
By taking the derivative of $\mathcal{L}$ with respect to $R_n^t$ and set it to zero, we have
\begin{equation} 
	\frac{\partial \mathcal{L}}{\partial R_n^t} = -\frac{\varepsilon_n^t L_n^s }{(R_n^t)^2} + \lambda = 0 \implies R_n^t = \sqrt{\frac{\varepsilon_n^t L_n^s }{\lambda}}. \label{eq:opt_cond}
\end{equation}
Then, we substitute (\ref{eq:opt_cond}) into $\sum_{n=1}^N R_n^t = R$ at optimum
\begin{equation}
	\sum_{n=1}^N \sqrt{\frac{\varepsilon_n^t L_n^s }{\lambda}} = R \implies 
	\lambda = \frac{ \left( \sum_{n=1}^N \sqrt{\varepsilon_n^t L_n^s} \right)^2}{R^2}.
\end{equation}
By substituting $\lambda$ back into (\ref{eq:opt_cond}), we obtain the optimal transmission rate allocation solution
\begin{equation} \label{tsa}
	R_n^{t,*} = \frac{R \sqrt{\varepsilon_n^t L_n^s}}{\sum_{m=1}^N \sqrt{\varepsilon_m^t L_m^s}}, \quad \forall n. 
\end{equation}
That is, the rate allocated to user $n$ is proportional to $\sqrt{\varepsilon_n^t L^s_n}$, i.e., the square root of the output feature dimension. 
Then, we substitute $R_n^{t,*}$ into (P1) and reduce the problem to the following pure CR optimization problem:  
\begin{subequations}
	\begin{align}
		\text{(P3):}	\quad  \max_{\boldsymbol{\varepsilon}^t} & \  \sum_{n=1}^{N}  Q_n^t  - \alpha \cdot \frac{BT}{NR} \left( \sum_{n=1}^{N} \sqrt{\varepsilon_n^t L_n^s}\right)^2 ,   \label{P3}  \\
		\text { s.t. } & Q_n^t  \ge Q_n^{\min}, \quad \forall n, \forall t,  \label{P3_c1} \\  
		&   \varepsilon_n^{\min} \le  \varepsilon_n^t \le   \varepsilon_n^{\max}, \quad \forall n. \label{P3_c2} 
	\end{align}
\end{subequations}

\begin{figure*}[!t]
	\centering
	\includegraphics[width=6.5in]{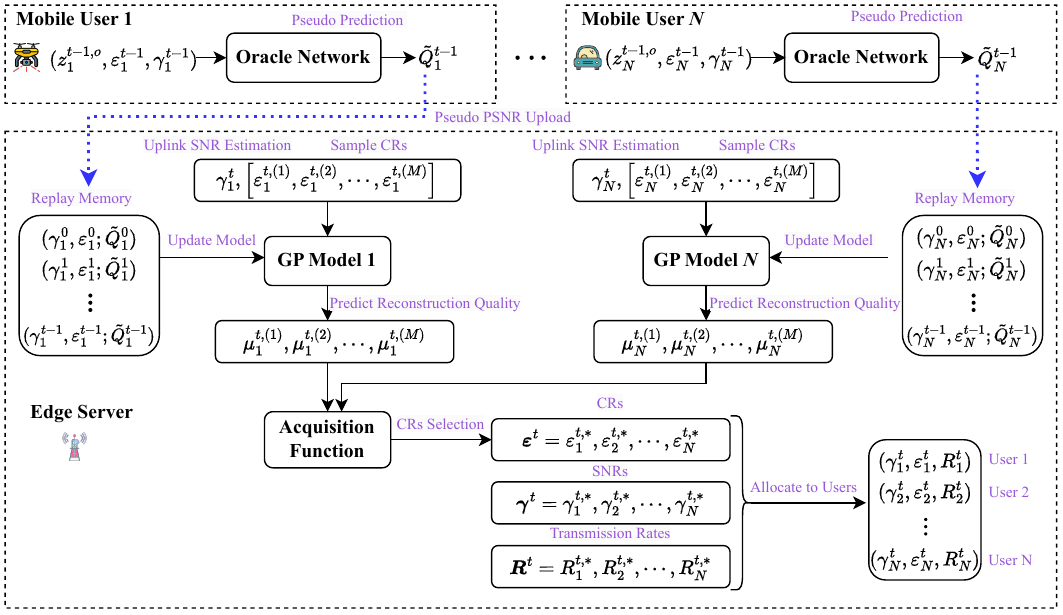}
	\caption{Proposed BO-based multi-user online CR selection method.
	}
	\label{fig3}
\end{figure*}

\subsection{BO-based CR Selection} \label{gp}
\subsubsection{GP modeling}
In section \ref{JSCC}, we have introduced an oracle network deployed on the user side to predict the reconstruction quality $\tilde{Q}_n^t$ as a function of $\boldsymbol{z}_n^t$, $\gamma_n^t$, and $\varepsilon_n^t$. In this case, we transform (P3)  as
\begin{subequations}
	\begin{align}
		\text{(P4):}	\quad  \max_{\boldsymbol{\varepsilon}^t} & \  \sum_{n=1}^{N}  \tilde{Q}_n^t  - \alpha \cdot \frac{BT}{NR} \left( \sum_{n=1}^{N} \sqrt{\varepsilon_n^t L_n^s}\right)^2 ,   \label{P4}  \\
		\text { s.t. } & \tilde{Q}_n^t  \ge Q_n^{\min}, \quad \forall n, \forall t,  \label{P4_c1} \\  
		&   \varepsilon_n^{\min} \le  \varepsilon_n^t \le   \varepsilon_n^{\max}, \quad \forall n. \label{P4_c2} 
	\end{align}
\end{subequations}

Still, (P4) is challenging to solve at the edge server side due to the following two reasons. First, the CR selection $\varepsilon_n^t$ is decided at the edge server, but the oracle network resides at the user side and relies on the local sensing data input $z_n^t$. As such, the edge server does not know $\tilde{Q}_n^t$ when making the CR selection decision. This is particularly challenging to maximize the objective in (\ref{P3}) and meanwhile satisfying constraint (\ref{P3_c1}). Second, the CR selections of the $N$ users are implicitly coupled by the limited transmission rate $R$. Therefore, the edge server needs to jointly solve the $N$ black-box optimization problems of all the users. 

To address these challenges, we propose a BO-based method as illustrated in Fig. \ref{fig3}. The framework comprises two key components, including a Gaussian process (GP) surrogate model and a CR decision acquisition function. Specifically, the GP surrogate model constructs a tractable mapping between $ \boldsymbol{v}_n^t= (\varepsilon_n^t, \gamma_n^t)$ and $\tilde{Q}_n^t$. In this way, we transform the original hard constraint $\tilde{Q}_n^t \ge Q_n^{\min}$ to a probabilistic constraint, enforced through the analytically tractable probability $P(\tilde{Q}_n^t \ge Q_n^{\min})$ obtained from the Gaussian marginal and its uncertainty quantification.
Then, we design an acquisition function that selects a CR decision to maximize the objective of (\ref{P3}) to achieve a trade-off between reconstruction quality and latency.

We denote the newest $t_u$ observations of user $n$ as $\boldsymbol{\mathcal{D}}_n^{t-1} = \left\lbrace \left( \boldsymbol{v}_n^{\tau}, y_n^{\tau} \right)  \right\rbrace_{\tau=t-t_u}^{t-1} $, where $y_n^{\tau}= \tilde{Q}_n^{\tau} + \tilde{\epsilon}_n^{\tau} $ is the noisy observation of reconstruction quality $\tilde{Q}_n^{\tau}$.  $\tilde{\epsilon}_n^{\tau} \sim \mathcal{N}(0, \tilde{\sigma}_n^2)$ is the observation noise and $\tilde{\sigma}_n$ is a learnable parameter which quantifies the uncertainly from the sample content variability, i.e., the reconstruction quality metric fluctuates due to difference between the transmitted samples even under the same $ \boldsymbol{v}_n^{\tau} $. 
Note that content dependence is handled by two complementary mechanisms. The observation noise $\tilde{\sigma}_n^2$ accounts for the within-distribution content variability under a given $\boldsymbol{v}_n^{\tau} = (\gamma_n^t, \varepsilon_n^t)$, whereas the online sliding-window update enables the GP mean to track the non-stationarity of the content distribution itself, as the most recent $t_u$ observations progressively replace stale ones.
We then suppose that $\tilde{Q}_n^{\tau}$ follows a Gaussian distribution with zero mean, i.e., $\tilde{Q}_n^{\tau} \sim \mathcal{GP}_n(0, k(\boldsymbol{v}, \boldsymbol{v}'))$, where $k(\cdot, \cdot)$ is a covariance function (kernel function) to describe the similarity of two input samples. Consequently, given the input $\boldsymbol{V}_n^{t-1}=\left[ \boldsymbol{v}_n^{t-t_u}, \cdots, \boldsymbol{v}_n^{t-1} \right]^{\top } $, the joint prior distribution of evaluations $\tilde{\boldsymbol{Q}}_n^{t-1}=\left[ Q_n^{t-t_u}, \cdots, Q_n^{t-1} \right]^{\top } $ can be written as 
\begin{equation} \label{prior_PDF}
	P\left( \tilde{\boldsymbol{Q}}_n^{t-1} \mid \boldsymbol{V}_n^{t-1} \right) = \mathcal{N}(\tilde{\boldsymbol{Q}}_n^{t-1}; \boldsymbol{0}_n^{t-1}, \boldsymbol{K}_n^{t-1}), \forall t. 
\end{equation}
Here, $\boldsymbol{K}_n^{t-1}$ is a covariance matrix with a size of $t_u\times t_u$, where the $(\tau, \tau')$th element is the value of kernel function of two input sample, defined as $[\boldsymbol{K}_n^{t-1}]_{\tau, \tau'}=k(\boldsymbol{v}_n^{\tau}, \boldsymbol{v}_n^{\tau '})$. The estimation of black-box function $\tilde{\boldsymbol{Q}}_n^{t-1}$ given the noisy observations $\boldsymbol{y}_n^{t-1} = \left[ y_n^{t-t_u}, \cdots, y_n^{{t-1}} \right]^{\top } $ follows a Gaussian conditional likelihood 
\begin{equation}
	P\left(\boldsymbol{y}_n^{t-1} \mid \tilde{\boldsymbol{A}}_n^{t-1},\boldsymbol{V}_n^{t-1} \right)=\mathcal{N}(\boldsymbol{y}_n^{t-1};\tilde{\boldsymbol{Q}}_n^{t-1}, \tilde{\sigma}_n^2 \boldsymbol{I}_n^{t-1}), \forall t,
\end{equation}
where $\boldsymbol{I}_n^t$ is an identity matrix. Then, with a new input $\boldsymbol{v}_n^{t}$, according to the prior PDF in (\ref{prior_PDF}) and Bayes' rule, we obtain the posterior PDF of $\tilde{\boldsymbol{Q}}_n^{t-1}$ as 
\begin{equation}
	P\left( \tilde{Q}_n^{t} \mid \boldsymbol{\mathcal{D}}_n^{t-1}  \right) =\mathcal{N}\left( \tilde{Q}_n^{t}; \mu_n(\boldsymbol{v}_n^{t}), \sigma_n^2(\boldsymbol{v}_n^{t}) \right).  
\end{equation}
Here, the mean and variance can be expressed as 
\begin{equation}
	\mu_n(\boldsymbol{v}_n^{t}) = \left[ \boldsymbol{k}_n^{t-1}(\boldsymbol{v}_n^{t}) \right] ^{\top} \left(\boldsymbol{K}_n^{t-1} + \tilde{\sigma}_n^2 \boldsymbol{I}_n^{t-1} \right)^{-1} \boldsymbol{y}_n^{t-1} ,
\end{equation}
\begin{align}
	&   \sigma_n^2(\boldsymbol{v}_n^{t}) = \notag  \\ & k(\boldsymbol{v}_n^{t},\boldsymbol{v}_n^{t})  
	- \left[ \boldsymbol{k}_n^{t-1}(\boldsymbol{v}_n^{t}) \right] ^{\top} \left(\boldsymbol{K}_n^{t-1} + \tilde{\sigma}_n^2 \boldsymbol{I}_n^{t-1} \right)^{-1}  \boldsymbol{k}_n^{t-1}(\boldsymbol{v}_n^{t}) ,
\end{align}
where $ \boldsymbol{k}_n^{t-1}(\boldsymbol{v}_n^{t}) = \left[ k(\boldsymbol{v}_n^{t-t_u},\boldsymbol{v}_n^{t}) , \cdots, k(\boldsymbol{v}_n^{t-1},\boldsymbol{v}_n^{t})  \right]^{\top} $. 

The prediction accuracy of the above GP highly relies on the design of the kernel function $k(\cdot , \cdot)$. Intuitively, there is a positive correlation between the input variables $(\gamma_n^{t}, \varepsilon_n^{t})$ and output $\tilde{Q}_n^{t}$, i.e., a larger $\gamma_n^{t}$ or $\varepsilon_n^{t}$ leads to higher $\tilde{Q}_n^{t}$. Considering both $\gamma_n^{t}$  and $\varepsilon_n^{t}$ are continuous variables, to improve optimization efficiency, we adopt the following polynomial kernel function
\begin{equation}
	k(\boldsymbol{v}, \boldsymbol{v}') = \psi_1 (\boldsymbol{v}^{\top} \cdot \boldsymbol{v}'+\psi_2),
\end{equation}
where $\psi_1$ and $\psi_2$ are learnable parameters of the kernel function updating in the following process of parameter update. 

\subsubsection{BO parameters updating}
Defining the set of parameters to be updated of the $n$th user as $\boldsymbol{\psi}_n =(\psi_{n, 1}, \psi_{n, 2}, \tilde{\sigma}_n)$, the update is performed by maximizing the log marginal likelihood 
\begin{align} \label{L_psi}
	\mathcal{L}(\boldsymbol{\psi}_n)  = & - \frac{1}{2}  (\boldsymbol{y}_n^{t-1})^{\top}\left(\boldsymbol{\Sigma}_n^{t-1} \right)^{-1} \boldsymbol{y}_n^{t-1} \notag \\ &- \frac{1}{2}  \log \left| \boldsymbol{\Sigma}_n^{t-1} \right| - \frac{1}{2}  \log 2\pi ,
\end{align}
where $\boldsymbol{\Sigma}_n^{t-1} = \boldsymbol{K}_n^{t-1} + \tilde{\sigma}_n^2 \boldsymbol{I}_n^{t-1} $. The first item in (\ref{L_psi}), which involves the observations, corresponds to the data-fit; the second reflects the complexity penalty; and the final one serves as a normalization constant. Accordingly, the gradient of $\mathcal{L}(\boldsymbol{\psi}_n) $ for the parameters $\boldsymbol{\psi}_n$ is 
\begin{align}
	\frac{\partial \mathcal{L}(\boldsymbol{\psi}_n) }{\partial \boldsymbol{\psi}_n} = &-  \frac{1}{2}  (\boldsymbol{y}_n^{t-1})^{\top}\left(\boldsymbol{\Sigma}_n^{t-1} \right)^{-1} \frac{\partial \boldsymbol{\Sigma}_n^{t-1}  }{\partial \boldsymbol{\psi}_n} \left(\boldsymbol{\Sigma}_n^{t-1} \right)^{-1}   \boldsymbol{y}_n^{t-1}  \notag \\
	& -  \frac{1}{2}\mathrm{tr} \left( \left(\boldsymbol{\Sigma}_n^{t-1} \right)^{-1} \frac{\partial \boldsymbol{\Sigma}_n^{t-1}  }{\partial \boldsymbol{\psi}_n}  \right) .
\end{align}
Finally, with the above gradient, the parameter update process for every $t_l$ indices is expressed as 
\begin{equation} \label{update_gp}
	\boldsymbol{\psi}_n \gets \eta \cdot \frac{\partial \mathcal{L}(\boldsymbol{\psi}_n) }{\partial \boldsymbol{\psi}_n},  
\end{equation}
where $ \eta$ is the learning rate.

\begin{algorithm}[t]
	\caption{Proposed BO-based Method for CR Selection and Resource Allocation}
	\label{bo_alg}
	\renewcommand{\algorithmicrequire}{\textbf{Initialization:}}
	\renewcommand{\algorithmicensure}{\textbf{Output:}}
	\begin{algorithmic}[1]
		\REQUIRE Observerd dataset $\boldsymbol{\mathcal{D}}^{0} = \left\lbrace \boldsymbol{\mathcal{D}}^{0}_n \right\rbrace_{n=1}^{N}$, parameters of GPs $\boldsymbol{\mathcal{\psi}}^{0} = \left\lbrace\boldsymbol{\mathcal{\psi}}_n^{0} \right\rbrace_{n=1}^{N}$
		\ENSURE Buffered dataset $\boldsymbol{\mathcal{D}}^{t} = \left\lbrace\boldsymbol{\mathcal{D}}^{t}_n \right\rbrace_{n=1}^{N}$, updated GPs $\boldsymbol{\mathcal{\psi}} = \left\lbrace\boldsymbol{\mathcal{\psi}}_n \right\rbrace_{n=1}^{N}$, selected CRs $\boldsymbol{\varepsilon}^{t} = \left\lbrace \varepsilon^{t}_n \right\rbrace_{n=1}^{N}$ and allocated rates $\boldsymbol{R}^{t} = \left\lbrace R^{t}_n \right\rbrace_{n=1}^{N}$.  
		\FOR{$t=0$ to $T_{\max}-1$}
			\STATE Obtain channel SNR $\boldsymbol{\gamma}^t$ of all users
			\STATE Select CR $\boldsymbol{\varepsilon}^t$ for each user \hfill $\triangleright$ Eq. (\ref{af}) 
			\STATE Allocate $\boldsymbol{R}^{t}$ for each user \hfill $\triangleright$ Eq. (\ref{tsa})
			\STATE Observe $\boldsymbol{y}^t$ from all users
			\STATE Store samples $\boldsymbol{\mathcal{D}}^{t} =\left\lbrace  \boldsymbol{\mathcal{D}}^{t-1}_n \cup (\boldsymbol{v}_n^t, y_n^t)\right\rbrace _{n=1}^{N}$
		\IF{$ t \mod t_l == 0 $}
			\FOR{$n=1$ to $N$} 
				\STATE Update GP   \hfill $\triangleright$ Eq. (\ref{L_psi}) to (\ref{update_gp})  
			\ENDFOR
		\ENDIF
		\ENDFOR
	\end{algorithmic}
\end{algorithm}

\subsubsection{Acquisition function design} \label{af_design}
At the index $t$, with the new observed SNRs $\boldsymbol{\gamma}^t$, we aim to seek the optimal $\boldsymbol{\varepsilon}^{t}$. 
With the GP, at the $t$, the probability that the performance metric $\tilde{Q}_n^{t}$ meets the constraint is denoted as 
\begin{equation} \label{CDF}
	P_n\left( \tilde{Q}_n^{t} \ge Q_n^{\min} \mid \boldsymbol{v}_n^{t} \right) =1- \Phi \left( \frac{Q_n^{\min}-\mu_n(\boldsymbol{v}_n^{t})}{\sigma_n(\boldsymbol{v}_n^{t})} \right), 
\end{equation}
where $\Phi$ is the cumulative distribution function (CDF) of the standard normal distribution. Then, we set a minimum confidence $c_n \in (0, 1)$ and require 
\begin{equation} \label{phi_c}
	P_n\left( \tilde{Q}_n^{t} \ge Q_n^{\min} \mid \gamma_n^{t}, \varepsilon_n^{t} \right) \ge c_n,
\end{equation}
which can be inferred as 
\begin{equation}
	\Phi \left( \frac{Q_n^{\min}-\mu_n(\boldsymbol{v}_n^{t})}{\sigma_n(\boldsymbol{v}_n^{t})} \right) \le 1- c_n.
\end{equation}
Then, by applying the inverse CDF of the standard normal distribution $\Phi ^{-1}$  to both sides simultaneously, we obtain 
\begin{equation}
	\frac{Q_n^{\min}-\mu_n(\boldsymbol{v}_n^{t})}{\sigma_n(\boldsymbol{v}_n^{t})} \le \Phi ^{-1}(1-c_n).
\end{equation}
Using the properties of $\Phi ^{-1}(1-x) = - \Phi ^{-1}(x)$, we rewrite the above equation as
\begin{equation}
	\mu_n(\boldsymbol{v}_n^{t}) - Q_n^{\min} \ge \Phi ^{-1}(c_n) \sigma_n(\boldsymbol{v}_n^{t}).
\end{equation}
For notation simplicity, we define $\beta_n=\Phi ^{-1}(c_n)$, which is actually a positive constant and depends on the requirement of $c_n$. Accordingly, the probabilistic constraint in (\ref{phi_c}) is transformed into a deterministic nonlinear constraint as
\begin{equation}
	\mu_n(\boldsymbol{v}_n^{t}) -  \beta_n  \sigma_n(\boldsymbol{v}_n^{t})\ge Q_n^{\min} .
\end{equation}
 
For all users, we denote the feasibility region as 
\begin{equation}
	\mathcal{F}=\left\lbrace \boldsymbol{\varepsilon}^{t} \mid \mu_n \left( \boldsymbol{v}_n^{t} \right) -  \beta_n \sigma_n \left( \boldsymbol{v}_n^{t} \right)  \ge Q_n^{\min}, \forall n  \right\rbrace. 
\end{equation} 
By using GP to surrogate $\tilde{Q}_n^{t}$, we aim to optimization the expectation of the GP, i.e., replace $\sum_{n=1}^{N}  \tilde{Q}_n^{t} $ in the optimization target to $\sum_{n=1}^{N}  \mu_n(\boldsymbol{v}_n^{t}) $. 
We then use the Monte Carlo sampling method to seek the optimal CR. Specifically, we generate $M$ independent samples from search space $\mathcal{S}=\left\lbrace \boldsymbol{\varepsilon}^{t, (m)} \right\rbrace _{m=1}^M$ with each $\varepsilon_n^{t, (m)} \in \boldsymbol{\varepsilon}^{t, (m)} $ uniformly sampled within $\left[ \varepsilon_n^{\min} , \varepsilon_n^{\max}  \right] $. Finally,  the optimal CRs $\boldsymbol{\varepsilon}^{t, *} $ are selected by solving the follow problem 
\begin{subequations}
\begin{align} \label{af}
\argmax_{ \ \ \ \boldsymbol{\varepsilon}^{t, (m)} \in \mathcal{S} } \   & \sum_{n=1}^{N}  \mu_n(\boldsymbol{v}_n^{t}) - \alpha \cdot \frac{BT}{NR} \left( \sum_{n=1}^{N} \sqrt{\varepsilon_n^{t} L_n^s}\right)^2, \\
	\text { s.t. } & \mu_n \left( \boldsymbol{v}_n^{t} \right) -  \beta_n \sigma_n \left( \boldsymbol{v}_n^{t} \right)  \ge Q_n^{\min}, \forall n.
\end{align}
\end{subequations}
The proposed BO-based CR selection and transmission rate allocation method is summarized in Algorithm \ref{bo_alg}.

\subsubsection{Computational complexity analysis} \label{cca}
The computational complexity of the proposed method consists of two parts. The first part is the update of the GP-based surrogate model using the observed dataset $\boldsymbol{\mathcal{D}}_n^{t-1}$. For a single user, the computational complexity is $\mathcal{O}(|\boldsymbol{\mathcal{D}}_n^{t-1}|^3)$, where $|\boldsymbol{\mathcal{D}}_n^{t-1}|$ denotes the cardinality of $\boldsymbol{\mathcal{D}}_n^{t-1}$. As introduced earlier, a fixed-size sliding window strategy is employed, which limits the number of samples used for model updates to at most $t_u$. Therefore, the computational complexity of the model update remains bounded by $\mathcal{O}(t_u^3)$, which is manageable and does not pose a computational burden.
The second part corresponds to the prediction stage, where the mean and variance must be computed for a total of $M$ sample points. Each prediction involves matrix inversion and polynomial dot product operations, resulting in a computational complexity of $\mathcal{O}(M t_u^2)$.
For the overall complexity of $N$ users, since we update GPs per $t_l$ indices, it is $\mathcal{O}(\sum_{n=1}^{N}M t_u^2)+\mathcal{O}(\frac{1}{t_l}\sum_{n=1}^{N}M t_u^3)$. This is determined by the update frequency $t_l$, window size $t_u$, and sample points $M$.

\section{Numerical Results} \label{result}
In this section, we evaluate the performance of the proposed multi-user online resource allocation framework and compare it with representative benchmarks.

\subsection{Experiment Setup}
\subsubsection{Dataset}
To simulate a heterogeneous multi-user semantic edge reconstruction environment, we employ four distinct high-resolution video datasets, each assigned to a separate user:
\begin{itemize}
	\item \textbf{BDD100K \cite{BDD100K}}: A large-scale, diverse driving video dataset with annotations for 10 tasks, supporting autonomous driving research\footnote{Given the scale of BDD100K, we select a subset of 50 videos for this study.};
	\item \textbf{MTDT \cite{MTDT}}: A multi-target detection and tracking dataset of UAVs, consisting of 50 HD video sequences captured from a delta-wing platform, with up to eight diverse UAV targets per sequence;
	\item \textbf{UAVDT-Benchmark-S (UBS) \cite{UB}}: A subset of the UAVDT dataset containing 50 urban videos captured by UAVs, designed for object detection and tracking tasks;
	\item \textbf{UAVDT-Benchmark-M (UBM) \cite{UB}}: Another subset of UAVDT, also comprising 50 videos, offering complementary scenarios to UBS.
\end{itemize}
Each dataset contains 50 videos, where 45 of them are used for training the JSCC encoder, 2 for validating, and 3 for testing the overall system performance. The three test videos from each dataset are further partitioned into three separate test groups, resulting in a total of three test datasets across all users. 
All videos are reformatted into image frames at 30 frames per second and resized to RGB images of dimension $3\times 512 \times 512$. We use the PSNR as the reconstruction performance metric for evaluation.

\subsubsection{Channel Condition}
We consider a Rayleigh fading channel model with the channel gain $h_n^t=\bar{h}_n \cdot \xi$, where $\bar{h}_n$ represents the average channel gain depended on the distance between the user $n$ and the edge server, and $\xi \sim \text{Exp}(1)$ is an random variable following independent exponential of unit mean. $\bar{h}_n$ follows the free-space path loss model 
\begin{equation}
	\bar{h}_n = G_A \left( \frac{3\times 10^8}{4 \pi f_c d_n} \right) ^{d_e}, 
\end{equation}
where $G_A$, $d_n$, and $d_e$ denote the antenna gain, the distance from the $n$th user to the edge server, and the path loss exponent, respectively. In this paper, we set $G_A=4.11$ and $d_e=3$ in the simulation. $f_c$ is the carrier frequency and set to 2.4GHz. 
In this work, we consider the edge server (ES) located at a fixed position with a height of 20 m. 
For the simulation environment, we consider a low-speed mobility scenario to ensure the quality of the captured image sources, which aligns with practical applications such as vehicular inspection and low-speed UAV aerial photography. Specifically, the velocity of the mobile users (i.e., vehicles and UAVs) is set to not exceed 5 km/h.
To evaluate the communication performance under different mobility scenarios, we generate the trajectories of the four users as follows:
\begin{itemize}
	\item User 1 is an autonomous vehicle moving at ground level along a rectangular trajectory centered around the ES, with a length of 100 m and a width of 50 m.
	\item User 2 is a drone capturing aerial imagery, flying at a constant altitude of 50 m along a square trajectory of 100 m $\times$ 100 m, also centered around the ES.
	\item User 3 and User 4 are drones tasked with capturing ground scenes, operating at heights of 20 m and 30 m, respectively. Both follow a rectangular trajectory of 100 m $\times$ 150 m centered at the ES.
\end{itemize}
Accordingly, the distance $d_n$ between the $n$th user and the edge server is computed as the Euclidean distance between the position of the user and the edge server. 

\begin{table}[!t]
		\centering
	\caption{Simulation Parameters of CR Selection}
	\begin{tabular}{|c|c|}
		\hline
		Weighting factor    & $\alpha=200$ \\ \hline
		Max samples      &  $t_u=20$ \\ \hline
		Constraints      & [33, 33, 26, 26] dB  \\ \hline
		Update frequency & $t_l=20$  \\ \hline
		Total rate       & $R=400$ Mbp/s  \\ \hline
		CR samples       &  $M=10000$  \\ \hline
	\end{tabular}
	\label{para_bo}
\end{table}

\begin{figure*}[!t]
	\centering
	\includegraphics[width=7in]{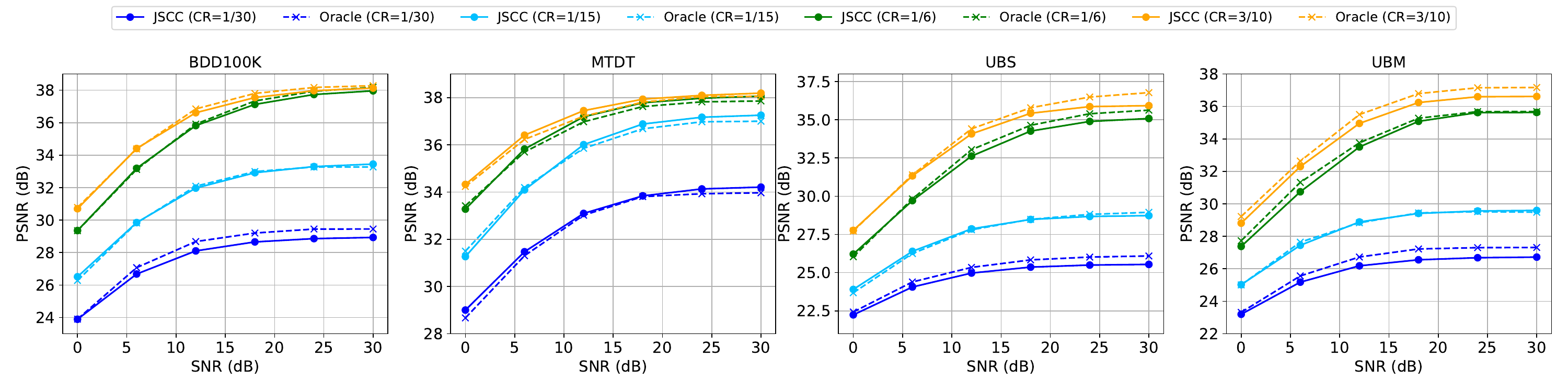}
	\caption{Results of JSCC and oracle network on 4 datasets.}
	\label{fig_jscc}
\end{figure*}

\subsubsection{Benchmarks}
For performance comparisons, we consider following benchmarks. 
\begin{itemize}
	\item \textbf{PSNR Maximum}: Using the highest available CR for all transmissions to maximize system reconstruction quality.
	\item \textbf{Latency Minimum}: Employing the lowest CR for all transmissions to minimize transmission latency. Notice that the scheme is likely to produce infeasible PSNR performance. 
	\item \textbf{PSNR Feasible}: Each user utilizes the oracle network to find the lowest CR that satisfies the individual reconstruction performance constraint.
	\item \textbf{DRL-SAC \cite{wang_SAC}}: A Soft Actor-Critic (SAC) algorithm is implemented as a representative state-of-the-art deep reinforcement learning benchmark. DRL-SAC is an off-policy actor-critic framework based on the maximum entropy principle, which encourages exploration and enhances robustness against environmental uncertainties during resource allocation.

\end{itemize}

\begin{table*}[!t]
	\centering
	\caption{Comprehensive Performance Comparison Across Methods (Results are averaged across three datasets over 20 independent trials.). Best results are highlighted in \textbf{bold} and second-best results are \underline{underlined}.}
	\setlength{\tabcolsep}{3pt}
	\renewcommand{\arraystretch}{1.15}
	\small
	
	\begin{tabular}{l|ccccc|ccccc|ccccc|c}
		\toprule
		\multirow{2}{*}{Method}
		& \multicolumn{5}{c|}{Constraint Satisfaction (\%) $\uparrow$}
		& \multicolumn{5}{c|}{PSNR (dB) $\uparrow$}
		& \multicolumn{5}{c|}{Latency (ms) $\downarrow$}
		& {Obj.} \\
		\cmidrule(lr){2-6} \cmidrule(lr){7-11} \cmidrule(lr){12-16} \cmidrule(lr){17-17}
		& User1 & User2 & User3 & User4 & Avg.
		& User1 & User2 & User3 & User4 & Avg.
		& User1 & User2 & User3 & User4 & Avg.
		& $\sum_{n=1}^N $ \\
		\midrule
		PSNR Max.
		& \textbf{99.11} & \textbf{98.67} & \textbf{100} & \textbf{100} & \textbf{99.45}
		& \textbf{40.34} & \textbf{36.69} & \textbf{35.78} & \textbf{33.69} & \textbf{36.63}
		& 150.99 & 150.99 & 150.99 & 150.99 & 150.99
		& 116.32\\
		Latency Min.
		& 56.33 & 44.22 & 80.78 & 38.11 & 54.86
		& 32.59 & 32.38 & 27.25 & 25.58 & 29.45
		& \textbf{16.78} & \textbf{16.78} & \textbf{16.78} & \textbf{16.78} & \textbf{16.78}
		& 114.44 \\
		PSNR Feas.
		& \underline{98.00} & \underline{98.55} & \underline{99.78} & \textbf{100} & \underline{99.08}
		& 34.25 & 34.27 & 27.62 & 26.91 & 30.76
		& \underline{26.72} & \underline{32.66} & \underline{22.82} & \underline{26.27} & \underline{27.12}
		& \underline{117.62} \\
		DRL-SAC
		& 95.33 & 93.22 & 97.78 & 97.03 & 95.84
		& \underline{38.03} & \underline{35.77} & 30.05 & 28.83 & 33.17
		& 83.53 & 78.85 & 83.27 & 84.83 &82.62
		& 116.16 \\
		\rowcolor{gray!15}
		Proposed
		& 98.45 & 97.11 & 98.55 & 98.00 & 98.03
		& 37.63 & 35.35 & \underline{32.34} & \underline{30.72} & \underline{34.01}
		& 82.41 & 78.43 & 84.64 & 85.49 & 82.74
		& \textbf{119.49} \\
		\bottomrule
	\end{tabular}%
	
	\label{tab_comprehensive}
\end{table*}

\subsubsection{Network Architecture and Training Setting}
The network architectures of both the JSCC and oracle models follow the same structural design as proposed in \cite{predict_JSCC}, and are applied to all users. We train the JSCC codecs and the oracle network for each user using their respective datasets. The batch size, number of training epochs, and learning rate are set to 4, 20, and $10^{-4}$, respectively. The training SNR is randomly sampled from the range [0, 30] dB. The training procedure for the oracle network is similar to that of the JSCC codecs. The extracted semantic features have a maximum dimension of $32 \times 128 \times 128$ (real-valued). A masking operation is applied to these features with a range from $\frac{1}{10}$ to $\frac{9}{10}$. After converting the real-valued features into complex-valued symbols for transmission, the resulting CR for each user $n$ falls within the interval $\left[ \varepsilon_n^{\min} = \frac{1}{30},\ \varepsilon_n^{\max} = \frac{3}{10} \right]_{n=1}^N$.
After training, the JSCC encoder and oracle network are deployed at the transmitter side (mobile user), while the JSCC decoder is deployed on the edge server.
Besides, the parameters used for updating the GP model are provided in Table \ref{para_bo}.

\subsection{Simulation Results of JSCC and Oracle Networks}
The results in Fig. \ref{fig_jscc} demonstrate the PSNR performance of JSCC and oracle prediction across four datasets under varying SNR conditions. As PSNR improves with an increasing SNR, the performance gain gradually diminishes in high-SNR regimes (SNR $\ge$ 20 dB), suggesting a saturation effect. For instance, with a CR of 1/30, the PSNR of the BDD100K dataset increases from 23.89 dB to 28.65 dB as the SNR varying from 0 dB to 18dB. However, when the SNR varying from 18 dB to 30 dB, the PSNR only increases from 28.65 dB to 28.93 dB. A similar saturation effect is also observed for the impact of CR. At CR=1/30, both JSCC and oracle achieve the lowest PSNR (24-28 dB at SNR=0 dB), indicating severe information loss under aggressive compression. 
Increasing CR to 1/6 yields the most significant improvement (4-10 dB across all datasets), demonstrating the importance of allocating sufficient transmission rate for feature preservation.
However, further increasing CR to 3/10 provides negligible gains (less than 1 dB), thus suggesting the opportunity to achieve an efficient accuracy-delay tradeoff by selecting a proper CR.

The figure also shows that the oracle network provides highly accurate PSNR predictions for the JSCC system. Across most CR and SNR configurations and all datasets, the oracle estimates closely match the true PSNR, with the maximum deviation being within 1 dB, even under low-CR settings. Therefore, in our semantic communication system, the pseudo-PSNRs produced by the oracle can be safely used to guide CR selection at the edge server. To further guarantee that the actual PSNR requirements are satisfied despite prediction uncertainty, we introduce a 1 dB safety margin in the PSNR constraints (e.g., a 30 dB requirement is enforced as 31 dB during CR optimization).

\begin{figure*}[!t]
	\centering
	\includegraphics[width=7in]{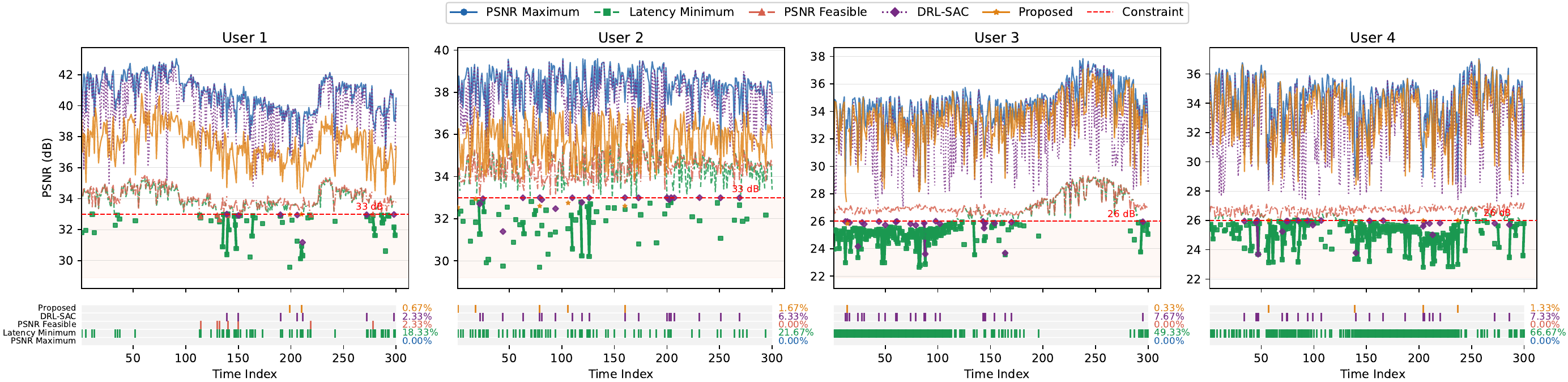}
	\caption{PSNR Performance and constraint violation rate of the proposed and benchmark methods across four users, evaluated on the first test dataset.}
	\label{fig_bo}
\end{figure*}

\begin{figure*}[!t]
	\centering
	\includegraphics[width=7in]{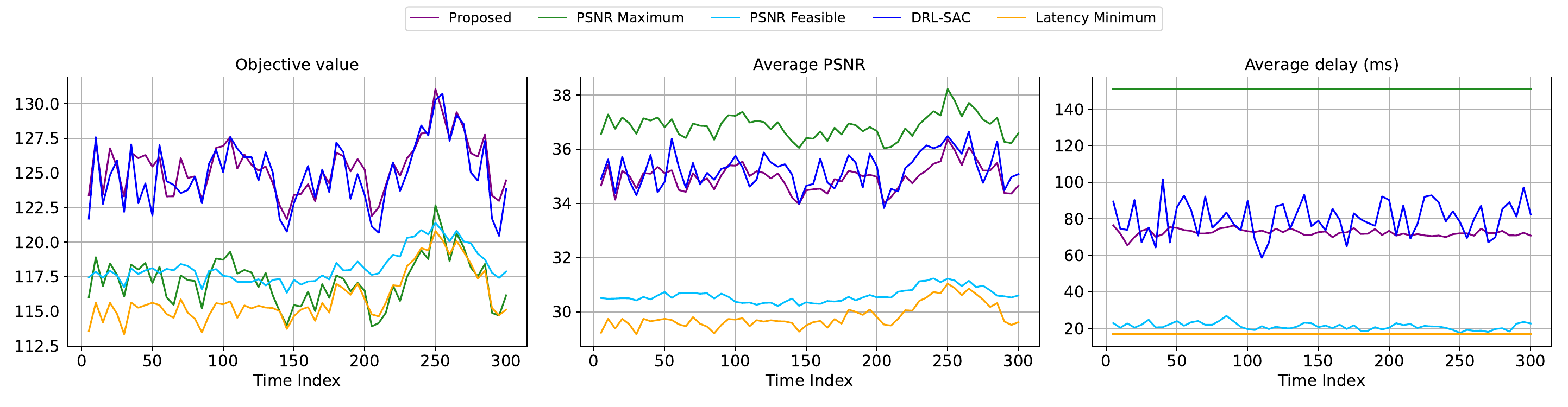}
	\caption{Performance as time evolves. Results are from the first test dataset and smoothed over 5-time-slot windows.}
	\label{fig_bo_s1}
\end{figure*}

\subsection{Performance Evaluation of the Proposed Method}

In this subsection, we evaluate the proposed BO-based online CR selection and resource allocation scheme and compare it with three benchmark methods using both per-user and system-level metrics. 
Table~\ref{tab_comprehensive} reports the constraint satisfaction rate (i.e., the portion of image indices in which the PSNR constraint $Q_n^{\min}$ is satisfied), average PSNR, average latency, and the overall objective value $\sum_{n=1}^{N} \left(  Q_n^t  - \alpha \cdot T_n^t/N \right)$ (i.e., the optimization target in (P1)) for each method across all users. 
A higher objective value indicates a better balance between reconstruction quality and latency.

As shown in the table, PSNR Maximum always selects the largest CR, achieving the highest constraint satisfaction ($\geq 98.67\%$ for all users) and the highest PSNR but incurring a prohibitive latency of $150.99$~ms per user. Conversely, Latency Minimum always uses the smallest CR, attaining the lowest latency of $16.78$~ms but violating the PSNR constraint in nearly half of all time slots (average satisfaction rate of only $54.86\%$). PSNR Feasible performs a per-user greedy search for the smallest feasible CR, yielding a high satisfaction rate ($99.08\%$) and low latency ($27.12$~ms) at the cost of substantially reduced PSNR ($30.76$~dB on average).

Among all the methods considered, the proposed BO method achieves the highest objective value of $\mathbf{119.49}$. Compared with DRL-SAC, the proposed method delivers a $2.19\%$ higher constraint satisfaction rate ($98.03\%$ vs.\ $95.84\%$) and a $0.84$~dB higher average PSNR ($34.01$~dB vs.\ $33.17$~dB) at comparable latency. Furthermore, compared with PSNR Maximum, the proposed method reduces latency by over $45\%$ while maintaining a satisfaction rate above $97\%$ for all users. Besides, it is crucial to clarify that approximately 2\% of constraint violations primarily arise from fundamental physical limitations, i.e., the stochastic wireless environment and neural codec characteristics. In other words, designing a scheme that strictly guarantees a 100\% satisfaction rate is practically impossible, and it is favorable to pursue an efficient tradeoff between reliability and latency.

To further illustrate the temporal behavior, Fig.~\ref{fig_bo} shows the time-varying PSNR of each user for the first test dataset (300 frames in each video sequence). 
For the proposed method, the PSNR trajectories remain consistently above the constraint for all users, with an average satisfaction rates of $99.33$\%, $98.33$\%, $99.67$\%, and $98.67$\% for Users 1--4, respectively. In comparison, the DRL-SAC method achieves satisfaction rates of $97.66$\%, $93.67$\%, $92.33$\%, and $92.67$\%, which are noticeably lower than the proposed method because conventional DRL struggles to enforce hard performance boundaries.
Although the PSNR Feasible method also satisfies the constraint almost all the time, its PSNR values stay only slightly above the threshold, leaving limited headroom for further performance improvement. As expected, PSNR Maximum achieves the highest PSNR but at the cost of very large latency, whereas Latency Minimum frequently violates the constraint, especially for Users~3 and~4.

We then analyze the system-level behavior of each method via the trajectories of the objective value, the average PSNR, and the average latency, as shown in Fig.~\ref{fig_bo_s1}. 
The proposed method achieves the highest average objective value (125.45), outperforming DRL-SAC (125.06), PSNR Feasible (118.19), PSNR Maximum (117.21), and Latency Minimum (115.99). This confirms that the BO-based optimization framework effectively exploits the trade-off between reconstruction quality and latency. In terms of average PSNR, PSNR Maximum yields the largest value (36.85 dB), followed by DRL-SAC (35.28 dB), the proposed method (34.98 dB), PSNR Feasible (30.61 dB), and Latency Minimum (29.84 dB). For average delay, Latency Minimum and PSNR Maximum result in constant delays of 16.78 ms and 150.99 ms, respectively. In contrast, the proposed method, DRL-SAC, and PSNR Feasible adapt the CR over time, leading to average delays of 72.42 ms, 80.23 ms, and 21.16 ms, respectively.

\subsection{Robustness Analysis under Varying PSNR Constraints}
To further evaluate the adaptability and robustness of the proposed method under different reconstruction performance requirements, we conduct additional experiments on the first test dataset with varying per-user PSNR constraints. While the original constraint vector is set to $[33, 33, 26, 26]$~dB for the four users, we tighten the constraints to $[34, 34, 26, 26]$, $[34, 34, 27, 27]$, $[35, 35, 27, 27]$, and $[35, 35, 28, 28]$, respectively, while keeping the total transmission rate $R$ unchanged.

\begin{figure}[!t]
	\centering
	\includegraphics[width=3.5in]{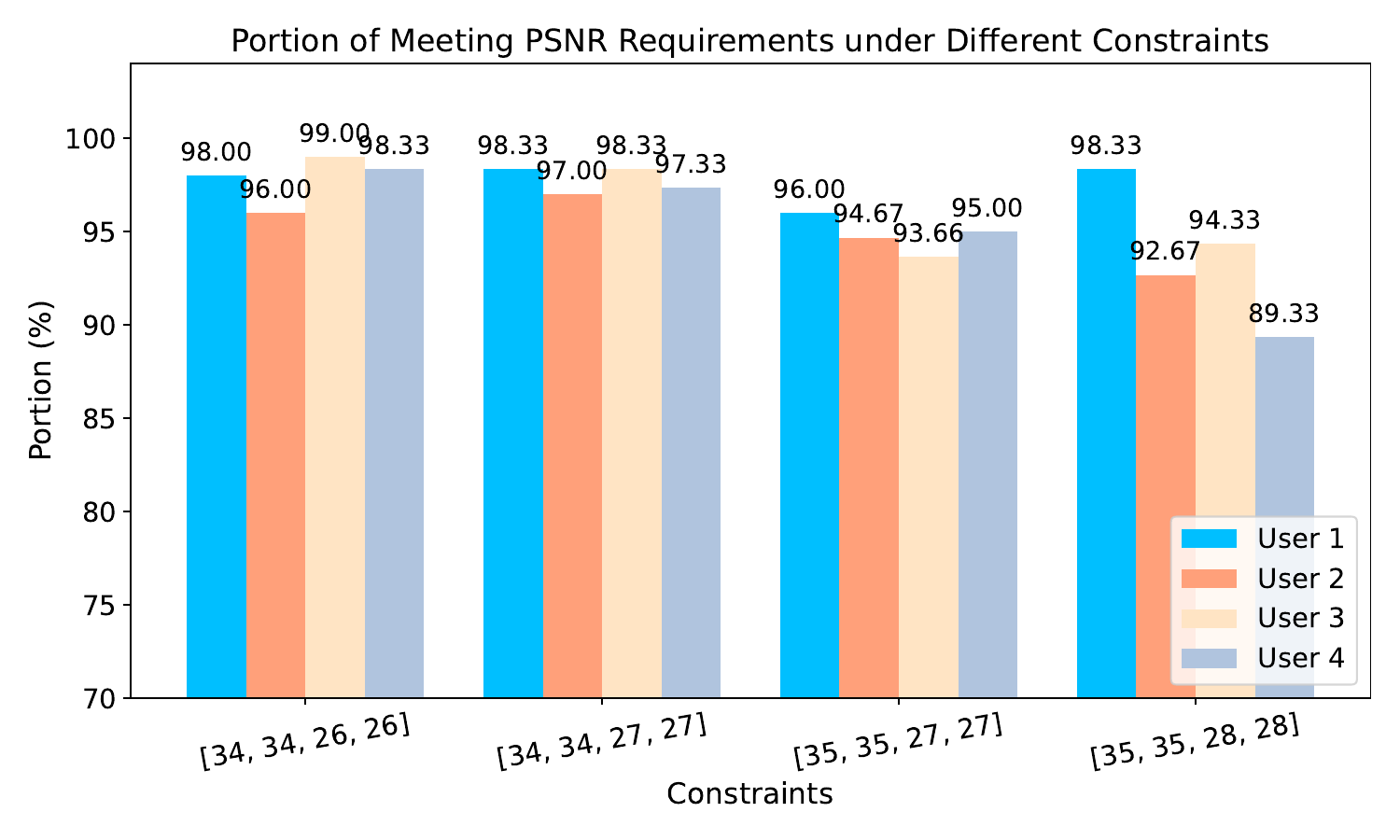}
	\caption{Portion of constraint satisfaction per user under different constraint settings.
	}
	\label{fig_cons}
\end{figure}

Fig.~\ref{fig_cons} shows, for each user and each constraint setting, the portion of time indices in which the PSNR constraint is satisfied. The proposed method maintains high satisfaction rates across all four settings. Specifically, the average satisfaction rates over time for Users~1--4 are [98\%,96\%,99\%,98.33\%], [98.33\%,97\%,98.33\%,97.33\%], [96\%,94.67\%,93.66\%,95\%], and [98.33\%,92.67\%,94.33\%,89.33\%] for the four tightened constraint vectors, respectively. These results indicate that the proposed method remains largely reliable under more demanding performance requirements: even under the most stringent constraints, the satisfaction probability for each user stays around or above $90\%$, with only a moderate degradation compared with the original setting.
Table~\ref{tab_lat2} reports the corresponding average latency per user for the different constraint sets. As the constraints become stricter, the latency values remain within a narrow range and do not exhibit a significant increase. For example, the average latency of User~1 is $83.03$~ms, $81.86$~ms, $81.37$~ms, and $81.99$~ms across the four settings. This shows that the proposed method adjusts the CR allocation efficiently in response to tighter performance requirements without substantially increasing transmission delay. Similar trends are observed for the other users, which confirms that the scheme effectively preserves the balance between reconstruction quality and latency under varying constraint levels.

Overall, the results in Fig.~\ref{fig_cons} and Table~\ref{tab_lat2} highlight the flexibility and robustness of the proposed method. It can adapt its resource allocation strategy to different PSNR requirements while maintaining high constraint satisfaction with nearly constant latency profiles. The slight decline in satisfaction rates under the most stringent constraints (e.g., User~4 dropping to $89.33\%$ in the fourth setting) is mainly due to the fixed total transmission resource $R$, and thus reflects the inherent tradeoff between resource availability and increasingly demanding performance targets. 

\begin{table}[!t] 
	\centering
	\caption{Average Latency Under Different Constraints}
	
	\begin{tabular}{lllll} 
		\toprule
		Constraints	&   User1     & User2      & User3     & User4    \\ \hline
		[34, 34, 26, 26]                & 83.03   & 77.03   & 86.15  & 85.73  \\ \hline
		[34, 34, 27, 27]               & 81.86   & 80.43   & 85.51   & 87.19     \\ \hline
		[35, 35, 27, 27]                & 81.37   & 75.42   & 85.76 & 88.76   \\ \hline
		[35, 35, 28, 28]                & 81.99    & 71.64   & 87.21   & 89.47    \\ \hline
		
		\bottomrule
	\end{tabular}%
	
	\label{tab_lat2}
\end{table}

\subsection{Sensitivity and Scalability Analysis}
\begin{figure}[!t]
	\centering
	\includegraphics[width=3.5in]{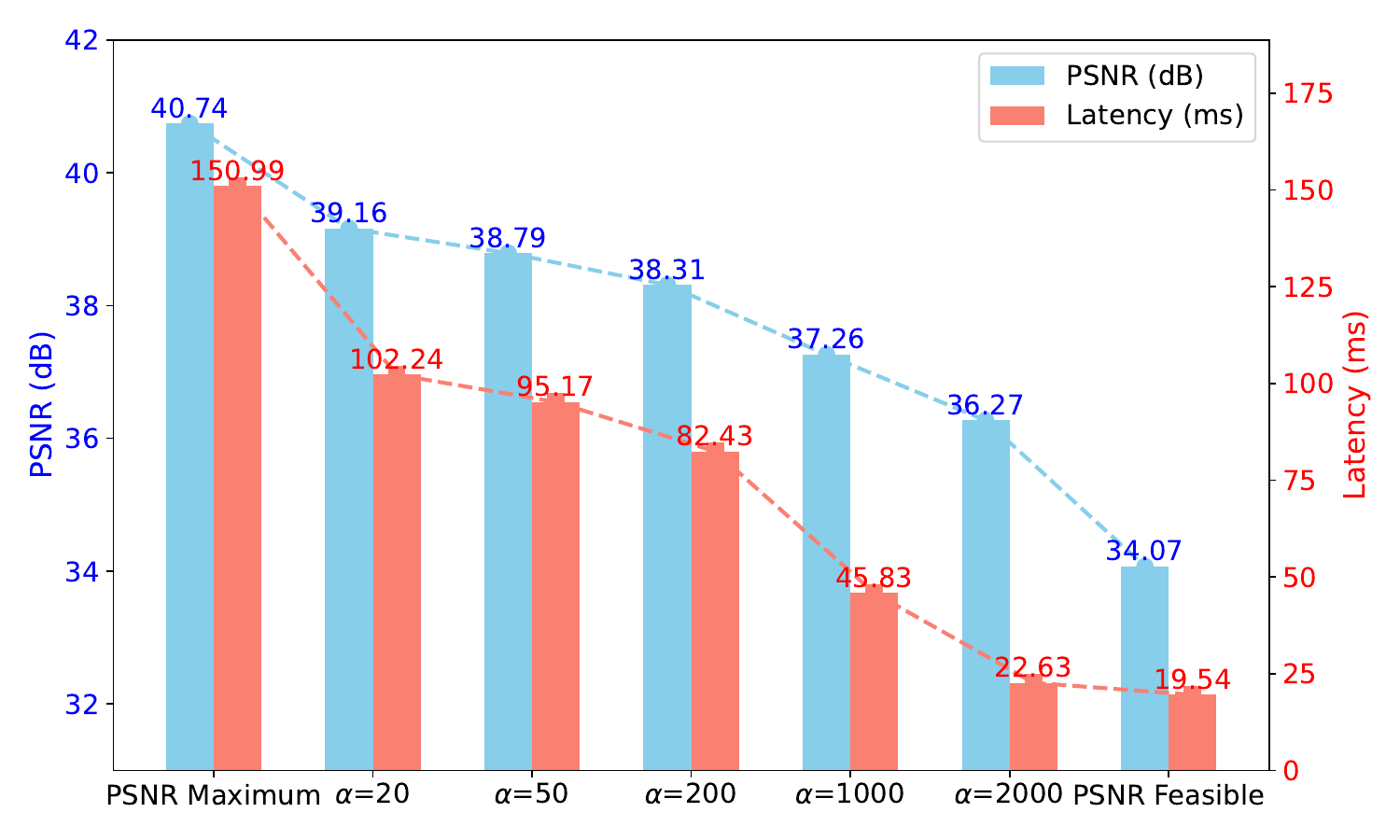}
	\caption{Influence of $\alpha$.
	}
	\label{fig_alpha}
\end{figure}

To investigate the flexibility of the proposed resource allocation framework, we analyze the impact of the weighting factor $\alpha$ on the trade-off between semantic performance and transmission latency. Fig. \ref{fig_alpha} presents a dual-axis plot illustrating the PSNR (left axis) and latency (right axis) for User 1 under different values of $\alpha$, with results averaged over the testing dataset. The compared strategies include PSNR Maximum, PSNR Feasible, and the proposed method with $\alpha=20, 50, 200, 1000, 2000$
As observed, when $\alpha$ is small (e.g., $\alpha=20$), the system prioritizes in enhancing reconstruction quality performance, yielding PSNR values close to those of the PSNR Maximum strategy (39.16 dB vs. 40.74 dB), albeit with higher latency (102.24 ms). As $\alpha$ increases, the system places greater emphasis on latency reduction. For instance, at $\alpha=500$, the latency drops to 22.63 ms, very near that of the PSNR Feasible method (19.54 ms), while the PSNR decreases to 36.27 dB. The default setting $\alpha=200$ strikes a balance, achieving a PSNR of 38.31 dB and a latency of 82.43 ms.
These results align with the formulation of (P3), where $\alpha$ controls the relative importance of latency in the objective function. Notably, the effect of varying $\alpha$ is equivalent to adjusting the available total transmission rate $R$ when $\alpha$ is fixed. This behavior demonstrates the system’s ability to adapt to different communication resource conditions: when transmission resource is abundant, it emphasizes performance; when resource is limited, it reduces latency while still satisfying performance constraints. 

\begin{table}[!t] 
	\centering
	\caption{Scalability Results under Different Users}
	
	\begin{tabular}{llll} 
		\toprule
			&   4 Users    & 8 Users      &  12 Users         \\ \hline
		Portion                & 98.03\%   & 97.21\%   & 96.11\%    \\ \hline
		PSNR (dB)              & 34.01    & 33.98   & 33.96        \\ \hline
		Latency (ms)            & 82.74   & 82.17   & 77.73    \\ \hline
		
		\bottomrule
	\end{tabular}%
	
	\label{tab_sca}
\end{table}

\begin{figure}[!t]
	\centering
	\includegraphics[width=3.5in]{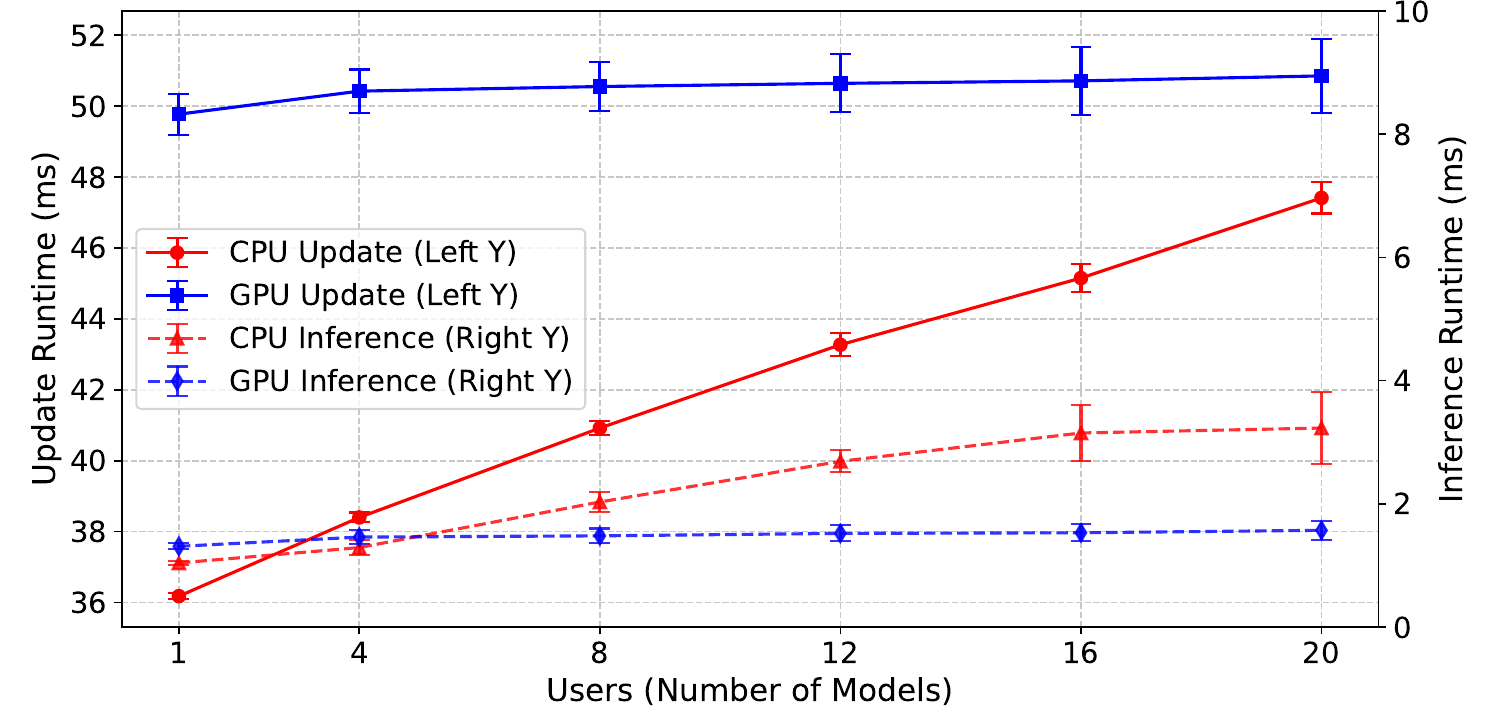}
	\caption{Model update and inference runtime under different users.}
	\label{fig_runtime}
\end{figure}

\begin{figure*}[!t]
	\centering
	\includegraphics[width=7in]{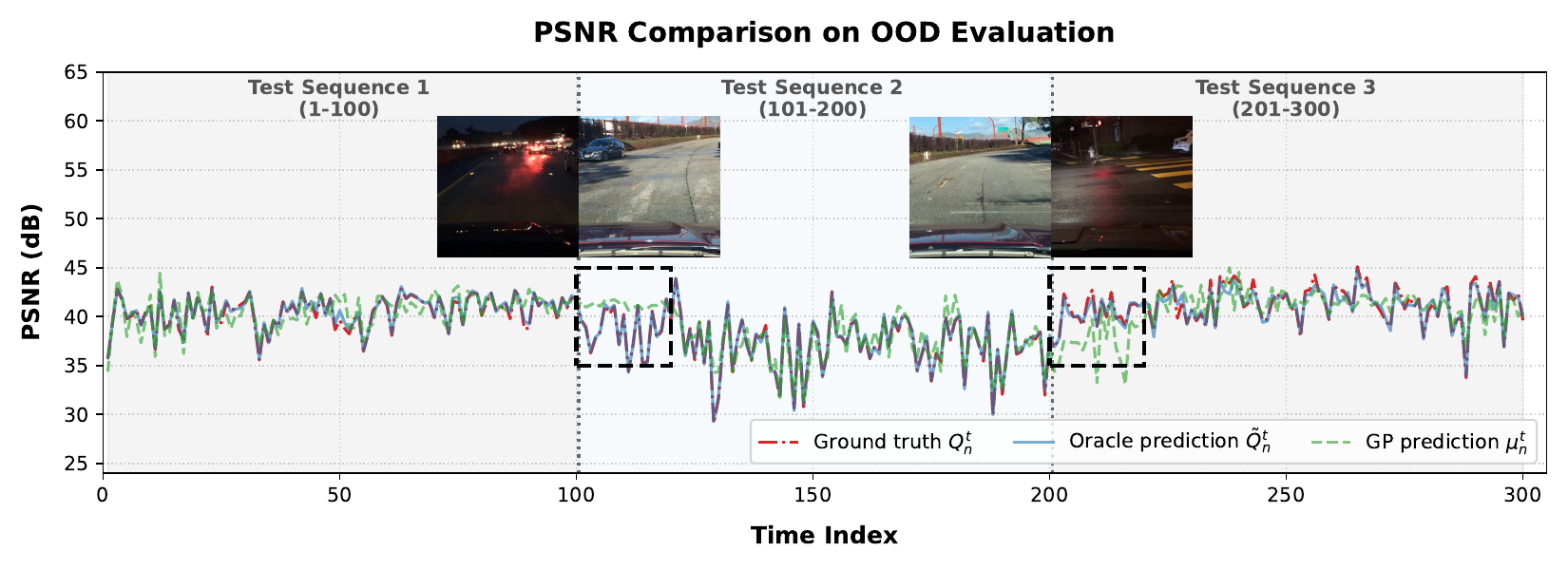}
	\caption{GP prediction under content distribution shift (OOD evaluation).}
	\label{fig_ood}
\end{figure*}

To evaluate the scalability of the proposed method, we extended the experiments to 8-user and 12-user scenarios. The 8-user case was constructed by incorporating a second testing dataset for users 5–8, with performance constraints set to [33,33,26,26,33,33,26,26]. The 12-user case further added the third dataset for users 9–12, with constraints [33,33,26,26,33,33,26,26,33,33,26,26]. Table V summarizes the average portion of constraint satisfaction, PSNR, and latency across all users.
The proposed method maintains high performance as the number of users increases: the satisfaction rate remains above 96\%, the average PSNR stays around 34 dB, and latency even slightly decreases due to more efficient multiplexing of resources. These results confirm the scalability of the framework to larger networks without significant degradation in system performance.

To validate the real-time feasibility of the proposed online resource allocation framework, we evaluate its empirical runtime overhead. The algorithm was implemented using GPyTorch to leverage parallel GP inference. The simulations were executed on an edge server equipped with an AMD Ryzen 7 9700X CPU and an NVIDIA RTX 5090 GPU. We vary the number of users from $N=1$ to $N=20$ and measured the execution time for both the model update phase and the inference (CR selection) phase. Fig. \ref{fig_runtime} illustrates the average runtime and standard deviation across different numbers of users. For the model update phase, the CPU runtime increases mildly from $36.18$ ms ($N=1$) to $47.41$ ms ($N=20$), which occurs periodically every tens of time slots and can be executed in parallel with the online inference process without incurring additional delay. More importantly, the inference phase—which is executed in every single time slot to allocate resources—incurs very small computational delay. When running on the CPU, the inference time increases from $1.04$ ms to $3.23$ ms. When running on the GPU, computing capability produces a nearly flat latency curve ranging from $1.31$ ms to $1.57$ ms. In practice, the $1.5$ ms decision-making overhead is negligible compared to the time slot duration of hundreds of milliseconds. This confirms that our proposed Constraint BO framework can safely satisfy the strict delay requirements of real-time edge inference systems.

\subsection{Robustness under Content Distribution Shift}
To evaluate the GP surrogate under an explicit distribution shift, we construct a controlled non-stationary a dedicated out-of-distribution (OOD) evaluation that concatenates three test sequences of BDD100K dataset with abrupt scene changes: a night scene (frames~1--100), a day scene (frames~101--200), and a night scene again (frames~201--300). The two scene transitions at frames~100 and~200 therefore emulate the day-to-night type of OOD content shift encountered in practical deployment. We record the ground-truth PSNR $Q_n^t$, the oracle prediction $\tilde{Q}_n^t$, and the GP prediction $\mu_n^t$ along the stream, with the GP updated every $t_l = 20$ slots. Fig.~\ref{fig_ood} reports the three PSNR trajectories. Two observations confirm the reliability of the implicit treatment under distribution shift. First, within each stationary segment the GP prediction $\mu_n^t$ closely tracks both the oracle prediction $\tilde{Q}_n^t$ and the ground-truth PSNR $Q_n^t$, despite using only $(\gamma_n^t, \varepsilon_n^t)$ as input; the abrupt drop in the absolute PSNR level of the day segment (frames~101--200) relative to the night segments is faithfully followed. Second, immediately after each scene change (the dashed boxes), the GP exhibits a brief transient deviation, because its sliding window still contains samples drawn from the previous distribution. Once the next update occurs (e.g., around frames~120 and~220), the GP rapidly re-learns the new content statistics and $\mu_n^t$ re-aligns with $Q_n^t$. The duration of this transient period is affected by the update interval $t_l$ and can be further shortened in practice by adopting a tighter update schedule (e.g., $t_l = 5$).

These results demonstrate that, owing to its online sliding-window adaptation, the proposed framework remains reliable under content distribution shifts, incurring only a short, bounded transient at the moment of an abrupt change.

\section{Conclusions and Future Work} \label{clu}
This paper investigated multi-user semantic communications focused on joint CR and transmission rate optimization. To address the challenge of unobservable reconstruction quality at the receiver, we developed a transmitter-side oracle network for quality prediction and a BO-based CR selection method. By employing GPs and a constraint-aware acquisition function, the proposed framework effectively balances reconstruction quality and latency under probabilistic performance guarantees. Experimental results on high-resolution video datasets demonstrate that our method achieves a 98.03\% constraint-satisfaction probability and reduces average latency by over 45\% compared to fixed-CR benchmarks.

We conclude the paper with some interesting working directions. First, the proposed method can be extended to more bandwidth-efficient access paradigms such as Non-Orthogonal Multiple Access (NOMA). In a NOMA-based semantic system, for instance, we can combine CR optimization with power-domain resource allocation.
Besides, the reliability of our method hinges on accurate prediction of the oracle network. To address potential performance degradation under OOD shifts—such as architectural changes or unseen environments—future work will integrate online calibration techniques like conformal prediction. These methods will dynamically adjust safety margins for probabilistic constraints, ensuring system robustness despite fluctuating oracle confidence.
Furthermore, while this paper focuses on continuous rate allocation and un-coded semantic transmission to demonstrate the core trade-off mechanism, incorporating practical wireless constraints—such as discrete Modulation and Coding Scheme (MCS) scheduling and advanced forward error correction (FEC) coding—into the constraint BO framework remains an important avenue for our future research.

\appendices

\ifCLASSOPTIONcaptionsoff
  \newpage
\fi

\end{document}